# Exploring Researchers' Interest in Stack Overflow: A Systematic Mapping Study and Quality Evaluation

Sarah Meldrum, Sherlock A. Licorish[*], Bastin Tony Roy Savarimuthu
Department of Information Science
University of Otago
Dunedin, New Zealand
{sarah.meldrum, sherlock.licorish, tony.savarimuthu}@otago.ac.nz

**ABSTRACT**

**Context**: Platforms such as Stack Overflow are available for software practitioners to solicit solutions to their challenges and knowledge needs. This community's practices have in recent times however triggered quality-related concerns. This is a noteworthy issue when considering that the Stack Overflow platform is used by numerous software developers. Academic research tends to provide validation for the practices and processes employed by Stack Overflow and other such forums. **Objective**: However, previous work did not review the scale of scientific attention that is given to this cause. Evidence resulting from such an analysis could be useful for understanding the focus of academic studies when considering Stack Overflow and how research is conducted on this forum. Furthermore, pertinent research quality evaluations may direct replication studies. **Methods**: Continuing from our preliminary work, we conducted a Systematic Mapping study involving 265 papers from six relevant databases to address this gap. In this work, we explored the level of academic interest Stack Overflow has generated, the publication venues that are targeted, the topics that are studied, approaches used, types of contributions and the quality of the publications that are written about Stack Overflow. **Results**: Outcomes show that Stack Overflow has attracted increasing research interest over the years, with topics relating to both community dynamics and human factors, and technical issues. In addition, research studies have been largely evaluative or proposed solutions; however, the latter approach tends to lack validation. The contributions of these studies are often techniques or answers to a specific problem. Evaluating the quality of all studies that were dedicated to software programming (58 papers), our outcomes show that on average only 58% of the developed quality criteria were met. **Conclusion**: Notwithstanding that research is continually aiming to understand Stack Overflow and other similar communities, further investigations are required to validate such studies and the solutions they propose.

**Keywords:** Stack Overflow; Crowdsourcing; Systematic Mapping Study; Quality Evaluation

* Corresponding author

# 1. INTRODUCTION

This work presents a systematic mapping study and quality evaluation of publications that are written about Stack Overflow. Stack Overflow (available via stackoverflow.com) is the leading Community Question and Answer (CQA) platform for programmers, with more than 10 million users, contributing some 16 million questions as of 2019[1]. This platform's success is evident through the fact that 92% of questions posted are answered, with a median answering time of 11 minutes [2]. Parnin and Treude [22] found that 84.4% of general web searches for jQuery API on Google resulted in at least one Stack Overflow outcome returned on the first page, showing that programmers can turn to Stack Overflow for questions already asked. Chen and Xing [6] also found data from Stack Overflow tags to be useful for creating an overall view of technologies trend.

Software engineering has a particular reliance on crowdsourced knowledge (and CQAs), given the community's general drive and emphasis on knowledge reuse [40]. CQA can benefit software practitioners seeking information, as it is likely that other practitioners have faced a similar problem, and so, a relevant question may have already been asked that has invoked a suitable answer. On the other hand, new questions may be created, and experts have the opportunity to lend their particular experience which allows them to solve a problem and gain the respect of their peers in the community. Beyond Stack Overflow, Yahoo!Answers[2], TopCoder[3] and Bountify[4] all provide crowd-based support for software practitioners' challenges. However, Stack Overflow tends to be the preferred choice for developers (e.g., [1]). While few would doubt the utility of Stack Overflow to software practitioners, questions regarding the quality of the responses generated to questions abound. For instance, Jin, et al. [13] studied how the need to 'win' reputation rewards can influence answer quality, or the drive and willingness to underestimate the quality of answers provided by others. Anand and Ravichandran [3] identified a need for separating answer quality and popularity, as the reward system (contributors on Stack Overflow are rewarded points for their contributions) provided by Stack Overflow can sometimes conflict with quality answers. These works suggest that there is need for caution when approaching Stack Overflow for recommended solutions to software engineering challenges.

However, Stack Overflow code reuse is ubiquitous. For instance, Abdalkareem, et al. [40] investigated code reused from Stack Overflow in mobile apps and found that 1.3% of the apps they sampled were constructed from Stack Overflow posts. They also discovered that mid-aged and older apps contained Stack Overflow code introduced later in their lifetime. An, et al. [41] also investigated Android apps and found that 62 out of 399 (15.5%) apps contained exact code clones; and of the 62 apps, 60 had potential license violations. In terms of Stack Overflow, they discovered that 1,226 posts contained code found in 68 apps. Furthermore, 126 Stack Overflow snippets were involved in code migration, where 12 cases of migration involved apps published under different licenses. Yang, et al. [42] noted that, in terms of Python projects, over 1% of code blocks in their token form exist in both GitHub and Stack Overflow. At an 80% similarity threshold, over 1.1% of code blocks in GitHub were similar to those in Stack Overflow, and 2% of Stack Overflow code blocks were similar to those in GitHub [42].

In fact, for many developers, online sources such as Stack Overflow are of utility when they are faced with issues that require knowledge they do not possess. This brings into question their likely understanding of such code, which in turn brings into question the quality of the software they produce. Furthermore, security complications may arise, as evidence has shown that Stack Overflow portal includes insecure code [43]. An example of how catastrophic code reuse could be is illustrated by Bi [44]. This author shows that a piece of Stack Overflow code was used in the NissanConnect EV mobile app, which accidentally displayed a piece of text reading "App explanation: the spirit of Stack Overflow is coders helping coders". This example illustrates that code reused from Stack Overflow and other similar portals are not always examined thoroughly. Although this example illustrates a non-threatening issue, many similar cases could introduce security and functionality-related problems if not inspected properly. Thus, it is important to investigate and understand the extent of code reuse occurring between software systems and online code resources such as Stack Overflow, but to also examine the studies themselves for appropriate quality so that the community is *confidently* aware of the state of play.

---

[1] https://en.wikipedia.org/wiki/Stack_Overflow

[2] https://answers.yahoo.com

[3] https://www.topcoder.com

[4] https://bountify.co

Rigorously designed academic work may help with validating the utility and reliability of Stack Overflow. However, previous work has given limited attention to the scale and critique of scientific research that is focussed on the Stack Overflow platform. This represents a shortcoming in terms of the research community understanding the focus of academic studies when considering Stack Overflow and how research is conducted on this forum. This study significantly extends prior work of Meldrum, et al. [20] which appears as a short paper (6 pages) in the proceedings of the EASE conference in 2017. This extension (with 75% additional content), covers five research questions which are considered in depth in this paper (which is 39 pages long). Of the five questions considered here, the short paper provided preliminary results for three questions. In that study [20], a preliminary investigation was conducted to understand the level of academic interest that is dedicated to Stack Overflow, the topics that are explored and how the studies are performed. In the current study we go one step further, and conduct a comprehensive investigation, focusing on the level of academic interest that is given to Stack Overflow, the topics that are researched, the approaches that are used and the type of contributions that are provided. In addition, we analyse the quality of all Stack Overflow studies that were dedicated to software programming (58 papers). The five research questions we answer are provided in Section 2.3.

In terms of contributions, this work provides a summary of the level of academic interest Stack Overflow has generated, the publication venues that are targeted, the topics that are studied, approaches used and types of contributions that are provided. In providing this summary we identify specific avenues for future research. Furthermore, pertinent research quality evaluations conducted in this work could direct replication studies, both in terms of the issues to consider and how studies should be designed. Finally, we adapted previous classification schemes that were used for studying research approaches, research contributions and research quality; the new schemes (protocols) created in this work are useful contributions to the software engineering community.

The remaining sections of this paper are organized as follows. We provide the study background and related work in the next section (Section 2). Thereafter, we provide our method in Section 3, before providing our results in Section 4. We then discuss our findings and their implications in Section 5, prior to considering threats to the study in Section 6. Finally, we provide concluding remarks and point to future research in Section 7.

## 2. BACKGROUND AND RELATED WORK

Crowdsourcing has become an effective way of solving problems, with individuals turning to the internet for help [1]. In particular, crowdsourcing knowledge allows individuals to seek the answer to any query, driven by their need for information. Within software engineering, Stack Overflow is available for software practitioners to solve their domain specific problems [9]. Unaware of research work validating this platform or the scale of evidence contributing to this cause, we survey the range of studies on Stack Overflow that are performed by academics. To conduct this study, the body of works covering crowdsourced knowledge, CQA platforms, software engineering and reviews focussed on these topics are reviewed in the rest of this section. In Section 2.1 crowdsourced knowledge and CQA are reviewed in general. This leads to Section 2.2 which reviews CQA portals in the context of software engineering. Finally, Section 2.3 examines similar reviews and revisits and situates the research questions posed in Section 1.

### 2.1 Crowdsourced Knowledge and CQA

Crowdsourcing involves using the power of the crowd (i.e., online communities) to solve a problem [12]. In particular, crowdsourced knowledge seeks to solve the problem of obtaining specific knowledge that is meant to reach a wide range of people, where individuals may work together or independently to solve problems [11]. Forms of crowdsourced knowledge include online platforms such as wikis (e.g., Wikipedia), forums and CQA platforms (e.g., Stack Overflow).

CQA platforms, such as Yahoo!Answers and Quora[5], provide a service where users can ask a question to find information they are looking for, and other users are able to respond with an answer by sharing their own knowledge or expertise [32]. This service benefits those who look to the internet to answer their questions or find particular information [27], as a CQA "is a platform for knowledge diffusion and propagation" [15]. In particular, if the information users are looking for is not found, then a CQA portal may allow them to post their own question(s) and gain the expertise of another user(s) at a later stage that they wouldn't have had otherwise.

---

[5] https://www.quora.com/

The success and popularity of CQA platforms are helped by users' need for both professional and personal information [39]. General CQA websites such as Yahoo!Answers, for example, contains over four hundred million questions with more than one hundred million members [21]. CQA platforms are heavily dependent on the participation of their users, and require effort from experts to find questions they are able to answer. Both those asking questions and those providing answers are deemed contributors, and are key to CQA platforms' success. To encourage user participation, CQA websites provide tools and employ gamification techniques [13], which serve to reduce the effort required by users [27]. These processes, while encouraging contributors' participation, can sometimes also invoke quality concerns, and particularly for software development practitioners [3]. We take a closer look at CQA and software engineering in the next subsection.

## 2.2 CQA and Software Engineering

Software engineering has a particular reliance on CQA portals; as software development problems are often specific to a particular context [22]. Thus, contextually driven CQA portals can be of utility to software development problem solving, and often support code reuse [40]. For instance, while CQA platforms such as the general Yahoo!Answers website are broad in topic scope, platforms such as Stack Overflow are domain-specific. These domain-specific CQA platforms can benefit software development practitioners seeking information, as it is likely that many other practitioners have faced similar problems, and so, relevant questions may have already been asked that have invoked suitable answers. On the other hand, new questions may be created, and experts are then able to lend their particular experience in the provision of answers which allows them to solve a problem and gain the respect of their peers in the community. It is noteworthy that Stack Overflow as a CQA website covers a range of software engineering topics. Such a targeted website, unlike the more general CQA portals (e.g., Quora), often include programming code as a solution to a problem (or part of a solution). Additionally, websites such as TopCoder and Bountify seek to provide code solutions only, sometimes with the requirement of a payment as a form of reward.

In terms of participation, Stack Overflow is the leading CQA platform for programmers [9]. This platform's success and users' (i.e., contributors) popularity are shown where 92% of questions posted on the Stack Overflow website are answered, and a median answering time of 11 minutes was reported in 2011 [19]. Parnin and Treude [22] found that 84.4% of web searches for jQuery API resulted in at least one Stack Overflow result on the first page, showing that programmers can turn to Stack Overflow for questions already asked. Cavusoglu, et al. [5] investigated the reward system of Stack Overflow, finding evidence that the gamification techniques used such as badges stimulated voluntary participation of users, and the reach of their knowledge could be widespread. Chen and Xing [6] found data from Stack Overflow tags to be useful in creating an overall view of technologies, which included the relationship among technologies and the trends they follow.

On the utility of the Stack Overflow platform itself for supporting software engineering and its potential implications, Ahmad and Ó Cinnéide [60] assessed GitHub projects which have reused Java code snippets, in terms of code cohesion over time. They found that 42% of the project classes exhibited reduced cohesion, most of which did not regain the cohesion exhibited prior to adding the Stack Overflow code snippet. Abdalkareem, et al. [61] measured the quality of projects by classifying code commits as bug fixing or non-bug fixing. They found that the percentage of bug fixing commits in each project file was larger after adding the Stack Overflow code snippets. Ragkhitwetsagul, et al. [62] conducted a survey of Stack Overflow users, which found that common issues associated with Stack Overflow answers include outdated solutions and buggy code. This was confirmed when identifying Java snippets included in open-source projects, 66% of which were considered to be outdated solutions and over 5% of which were considered to be buggy. Campos, et al. [63] extracted JavaScript snippets from Stack Overflow and analysed them in terms of code rule violations (using ESLinter), with the most common violation types relating to style issues (82.9% of the violations). A small number of the code snippets associated with violations were found to be used in GitHub projects. Nikolaidis, et al. [64] measured quality in terms of technical debt (the effort required to fix code inefficiencies), determining that the Java code snippets were actually associated with an overall lower technical debt density than the project code. However, there were a number of cases when the code snippets were associated with a much larger technical debt density than the project code.

Other studies have examined Stack Overflow code snippets for characteristics indirectly related to code quality. Treude and Robillard [65] assessed whether Java code snippets on Stack Overflow are self-explanatory. The answer text and code comments were removed from a sample of code snippets. They were presented to GitHub users, who determined that less than half of the code snippets were considered to be self-explanatory. The inappropriate use of a Stack Overflow snippet in a software development project, due to the snippet not being self-explanatory, is likely to reduce the quality of the project code. Therefore, examining the question and answer (including the user comments) associated with the code snippet may be needed in order for it to be self-explanatory. Wu, et al. [66] examined how Stack Overflow code snippets were used in open-source projects, finding that 44% of the files containing Stack Overflow snippets were modified prior to use in the file. This may be an indication of a lack of code quality associated with the original snippet. A survey of developers (who use Stack Overflow) found that 32% of respondents re-implemented code snippets rather than reuse them in their original state because of the perception that such code snippets may be of poor quality. Yang, et al. [67] also found that the exact duplication of code snippets was rare in the Python code projects examined.

Other human-related and gamification curiosities are also surfacing in academic research. For instance, Ginsca and Popescu [8] investigated the relationship between Stack Overflow user profiles and answer quality, finding correlations between a more complete user profile and answer quality. Jin, et al. [13] studied how the need to 'win' reputation rewards can influence answer quality, or the drive and willingness to underestimate the quality of answers provided by others. Anand and Ravichandran [3] identified a need for separating answer quality and popularity, as the reward system provided by Stack Overflow can sometimes conflict with quality answers. Additionally, studies such as Treude, et al. [35] and Asaduzzaman, et al. [4] have investigated how the quality of the question itself can affect the quality of the answer that is provided. These works all point to the need to be vigilant when approaching CQA websites such as Stack Overflow for recommended solutions to software engineering challenges. Therein lies the opportunity for academic research to aid with validation of the content on this portal (e.g., see [8] for an assessment of CQAs more generally), however, such works must be rigorously executed in promoting the software engineering community's trust. We examine the studies that have been dedicated to the review of this state of play next.

## 2.3 Related Reviews and Research Questions

There have been some related reviews in the domain of CQA portals. Shah, et al. [29] developed a research agenda for social Q&A portals, identifying three challenges for research including data collection, data analysis and evaluation. Shah, et al. [28] provided a framework to understand the online environment of Q&A portals, consisting of modalities (services), motivations (users), and materials (content). Srba and Bielikova [32] completed a survey on the general CQA domain to provide researchers with an overview of literature and theories, finding two key areas for future research – preservation of long-term sustainability and transferability of CQA systems.

Examining the findings of these studies further, it is noted that previous researchers tend to use terminology inconsistently, both with multiple terms for a particular idea and some terms referring to multiple ideas. For instance, in the search for works focussed on CQA, studies have used terms such as "question routing" as a replacement for "expert finding" [32]. This indicates a need for further academic work to help with validation. In fact, while these two reviews focus on CQA in a general context, we were unable to find a review in the field of software engineering. Our work addresses this gap left by prior work ([28], [29] and [32]) by focussing on the use of CQA portals particularly focussing on Stack Overflow, a portal that is routinely used by the software engineering community. Our work also significantly extends the preliminary investigations of Meldrum, et al. [20] as indicated in the Introduction section. We have developed five research questions to investigate and understand CQA portals in the context of software engineering, and have thus focussed on Stack Overflow as a platform given its widespread use by software development practitioners and utility (refer to Section 1). We first created **RQ1**. *(a) What level of academic interest has the Stack Overflow platform generated over time, and (b) what publication venues are targeted?*, to gather insight into the academic interest into Stack Overflow and where researchers target for publication. For example, we are interested in knowing whether academic interest on Stack Overflow has increased over time and also whether researchers target conferences and journals equally. We next designed **RQ2**. *What* *software engineering topics are most frequently explored on Stack Overflow?*, which allows us to investigate themes of the overall topics that are investigated by academic researchers focussed on Stack Overflow. Answering this question, in particular, will assist in identifying gaps in research areas that need specific attention. Subsequently, we created **RQ3**. *What research approach(es) are used to investigate these topics?*, which seeks to understand the type of research studies (e.g., platform evaluations or developed solutions) that are performed on Stack Overflow, and particularly in terms of the utility of the approaches that are used. For example, we are interested in scrutinising to what extent the research outputs provided *validated solutions* when compared to *proposed solutions* (i.e., where the solutions were not validated). The next question was designed, **RQ4**. *What form did the contribution of the research take?*, to investigate the type of contribution or output of research studies that are dedicated to Stack Overflow. For example, we are interested in understanding if researchers are focussed on developing tools to enhance Stack Overflow knowledge or just investigating the type of content that is provided. Finally, we designed **RQ5**. *What is the quality of research conducted on Stack Overflow?*, to understand the quality of the research that is focussed on Stack Overflow, and particularly those studies that investigate aspects related to actual software development and programming (i.e., as against say "contributors' profile").

As contributions, we anticipate that evidence resulting from our analyses could ascertain the landscape of academic interest Stack Overflow has generated, the publication venues that are targeted, the topics that are studied, approaches used and types of contributions that are provided. Through this analysis the community could gain confidence in the oversight that is provided by the academic community and identify specific avenues for future research. Furthermore, none of the prior works have considered quality evaluations of work conducted on Stack Overflow. Thus, our results on research quality evaluations provides directions to researchers both in terms of the issues to consider when studying Stack Overflow and how studies should be designed. Our findings can also be used to direct future replication studies. We examine the approaches that were used to design and implement our study next.

## 3. METHOD

In this section we discuss the method and techniques that were followed, aligned with the systematic mapping process and quality evaluation, to answer the five research questions outlined in Section 2. In Section 3.1 we detail the protocol development at the core of the systematic mapping process. In Sections 3.2 to 3.4 we detail the processes we used to conduct the systematic mapping study, including details on the search strategy, selection criteria and final process. Section 3.5 outlines the classification schemes that were developed for this study. The steps outlined in Sections 3.2 to 3.5 help us to answer RQ1, RQ2, RQ3 and RQ4. Finally, Section 3.6 presents the process that facilitated the assessment of paper quality, in order to answer RQ5.

### 3.1 Protocol Development and Overview

This study follows a Systematic Mapping process, which provides a way for understanding the "structure of the type of research reports and results that have been published by categorizing them" [25]. A systematic map can be useful for trying to define unknown research areas, and can also visualise these. A systematic map is similar to a systematic review in that it can identify research gaps, however it is different in that a map aims to establish what current research exists by looking at the overview of the studies, while systematic reviews seek to look at studies in more depth, meaning a narrower focus is often taken [25]. The study takes on both perspectives, with the first four research questions (RQ1-RQ4) aimed to providing details wider in scope, and the last question (RQ5) providing much deeper meaning.

A systematic map is appropriate for this study as it gives an overview of the research that has been done, which in turn, helps to identify the gaps in that same area of research [52]. These can be identified by visualising the map as a two-dimensional bubble plot which shows the proportion of research works that have been done in certain areas relative to others. In this study we follow the guidelines of Peterson, et al. [25], as demonstrated in Figure 1. For each process (shown in the top row of the diagram), there is an outcome (shown in the bottom row), which ultimately results in the final systematic map. However, there was need at times to adopt aspects of the systematic review process, moving beyond "keywording using abstracts". When we could not satisfactorily extract the appropriate content from abstracts we have reviewed the paper [14]. Additionally, we conducted deeper evaluation and analysis to shed light on the quality of Stack Overflow research studies (58 papers) that have focussed on software development and programming (refer to Section 3.6).

## 3.2 Search Strategy and Piloting

Having defined the five research questions, we then conducted our search. We searched for the terms 'Stack Overflow' and 'StackOverflow' in the titles and abstracts of articles on IEEE Xplore Digital Library, ACM Digital Library, ScienceDirect, Scopus, Wiley InterScience and Springer Database, in line with the databases recommended for conducting reviews and systematic mapping studies [7, 53]. These search terms were created as recommended by earlier studies that have focussed more widely on CQA (e.g., [32]), where "Stack Overflow" was noted to be specific to the Stack Exchange crowd-based portal which offers support for software practitioners' challenges. We initially included Google Scholar which returned around 7,000 search results. However, searches done by the first two authors of the first five pages of the 7,000 results returned showed that most of these results were irrelevant, largely focussing on blog articles on Stack Overflow or the memory-related concept "buffer overflow". Of the 50 articles explored for the five pages (i.e., 10 articles for each page), 17 were blog related, 19 were about "buffer overflow" and the rest covered contents related to Stack Overflow academic research which were covered by the aforementioned databases. We only checked the first five pages of Google Scholar's returned results as 96% of users only look at the first page when searching [54], and thus, there was a perception that five pages would go beyond the typical search behaviour of normal users in providing us exhaustive coverage of the search results that were returned. Given the abovementioned evidence that most of the results on these five pages were not relevant, we excluded the Google Scholar database from our study.

Our searches were done in January 2017, and were restricted to "from 2009 to 2017" due to Stack Overflow being created at the end of 2008 (roughly eight years of Stack Overflow studies were surveyed). Further, we looked to exclude results that contained 'buffer' in the title or abstract due to conflict with the computer memory management concept, given that a random check of 10 returned studies from five of the six databases (only five studies were returned by Wiley Interscience) revealed that all studies with this word in the title were related to operating systems memory management [51]. In fact, more specifically, our checks of these results confirmed that the excluded papers considered protection against stack (buffer) overflows and related memory issues that can cause security vulnerabilities, indeed confirming that these papers were not relevant to our study. As a second layer of checks before omitting all of the returned results that contained 'buffer' in the title or abstract we closely examined all the results from IEEE Xplore that included 'buffer' (i.e., 16 titles and 32 abstracts), where we observed that they all referred to the irrelevant computer memory management concept. Search results were thus limited to those which did not include 'buffer' in the title or abstract. A summary of the search results returned is shown in Table 1. The results from the searches were saved as a comma separated values (CSV) file to facilitate our analyses.

**Table 1. Database searches**

| Database | Stack Overflow | StackOverflow | Total Studies |
| --- | --- | --- | --- |
| IEEE Xplore Digital Library | 124 | 54 | 178 |
| ACM Digital Library | 105 | 56 | 161 |
| ScienceDirect | 10 | 4 | 14 |
| Scopus | 255 | 118 | 373 |
| Wiley Interscience | 5 | 0 | 5 |
| Springer Database | 205 | 148 | 353 |
| ∑ | ***704*** | ***380*** | ***1,084*** |

## 3.3 Selection (Inclusion/Exclusion) Criteria

From the total 1,084 papers that were retrieved from the searches, 383 were duplicates, as determined from the title and abstracts, leaving 701 papers remaining. These papers were further screened for relevance by checking against our inclusion/exclusion criteria (see below). We first checked the title and abstract against our criteria. If there was still uncertainty about relevance then the paper itself was reviewed. Our full criteria are as follows:

Include paper if:

- From 2009 (Stack Overflow was created in September 2008) – this was established through our initial search query.
- 'Contributes to the body of knowledge' on Stack Overflow such as papers that investigate features/characteristics of Stack Overflow, or use the data from Stack Overflow. This means that the papers do not just describe Stack Overflow.

Exclude paper if:
- Not in English.
- Used as an example only (i.e., paper doesn't discuss Stack Overflow itself).
- Not peer-reviewed. This could include summaries, conference proceedings, or secondary studies.
- Refers to computer memory management 'Stack Overflow', as opposed to the online platform.
- It is a duplicate of another paper already included.

Through the screening process we found 14 proceedings (as part of the 701 papers cohort) which were all checked for relevant papers. Checks were done using a snowballing approach, where the titles and abstract of all papers in the 14 proceedings were perused for potential relevance to our study [50]. We found one new paper (though there were numerous duplicates), resulting in 702 papers altogether that were checked against our inclusion/exclusion criteria.

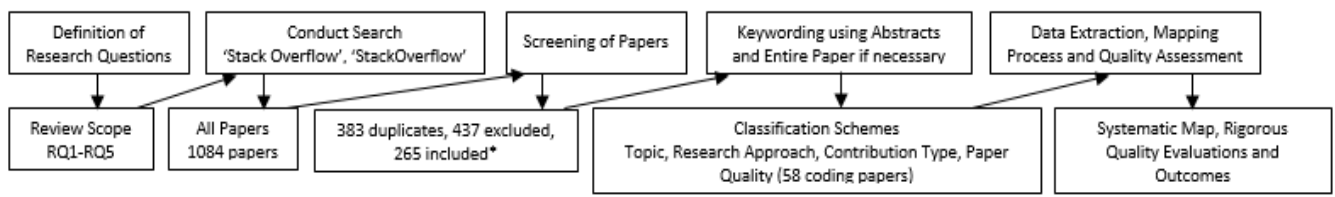

*One additional paper was added from the conference proceedings checks

**Figure 1. Systematic mapping process**

## 3.4 Final Review Process

The selection criteria excluded 437 papers and included 265 for further analysis, as detailed in Figure 1. To make sure that we had included an exhaustive list of papers, informal reliability checks were done between the three authors where 50 random papers were checked from each list (included and excluded). All three authors reviewed the titles and abstracts of these papers independently before then providing an indication of whether or not the paper should be included or excluded. These choices were then checked for consistency, revealing full agreement on all included and excluded papers (i.e., 100% agreement). The 265 papers included were then classified largely by the first and second authors, initially by reading the abstract, and if required looking at the paper itself. These were recorded in a Microsoft Excel spreadsheet where relevant information were extracted to answer our research questions (see the final list of included papers in Appendix A).

To ensure the delivery of reliable and valid outcomes, reliability checks were conducted by the first and second authors for verifying topics (RQ2), research approaches (RQ3), contribution types (RQ4) and paper quality (RQ5). While it was straightforward to determine the number of publications and where these were published in answering RQ1, and the classification schemes used for answering RQ3 and RQ4 were succinct, and thus, facilitated total agreement (of 50 random papers categorised and compared), deriving the topics in answering RQ2 required that we implement steps to facilitate a shared understanding. We initially agreed that topics would be coded using 2-4 relevant keywords based on the focus of the work, and the first and second authors coded our sample of 50 randomly selected papers independently. Thereafter, we reviewed our codes to examine the level of overlap, and noticed that in all instances our choices duplicated at least one keyword. As a result, we inferred that there was indeed shared understanding, and so, the first author's codes were retained and the topics for all remaining papers were coded. Thereafter, topics that emerge were compared using a constant comparison approach against previously identified topics [59], and appropriate adjustments were made leading to our specific pool of high level topics. Finally, these classifications were taken and used to create the final systematic maps – these are discussed as part of the results in Section 4. The quality evaluation to answer RQ5 required much more effort, as the entire paper needed to be reviewed deeply against a developed classification scheme. This aspect is considered in Section 3.6.

## 3.5 Classification Schemes

We created an exhaustive classification scheme in Table 2 following the guidelines of Wieringa, et al. [37], Shaw [30] and Paternoster, et al. [23]. Of note is that Wieringa, et al. [37] focused on research approaches (or types) while Shaw [30] and Paternoster, et al. [23] focused on classifying research contribution. An initial classification of 50 abstracts for all three aspects was completed to ensure the schemes were suitable for classifying the Stack Overflow papers. Small changes were then made to refine these, before the full classification was completed. For instance, when developing the classification scheme for studying research approach in Table 2, the *Opinion* and *Personal Experience* categories were removed from Wieringa, et al.'s [37] original classification scheme as our inclusion/exclusion criteria eliminated such studies. In addition, the *Solution/Validation* category was combined, as there were no papers in our sample that were exclusively validation papers. Furthermore, in terms of the classification scheme for evaluating research contribution in Table 2 (inspired by Shaw [30] and Paternoster, et al. [23]), we removed the *Analytical Model* category, and expanded the *Solution/Answer* category. These outcomes were then used to answer RQ3 and RQ4.

**Table 2. Classification schemes for answering RQ3 and RQ4**

| | |
|---|---|
| **Research Approach** (based on Wieringa, et al. [37]) | |
| Evaluation Research | Papers that evaluate the current state of play, such as the state of Stack Overflow or particular techniques (e.g., evaluation of Stack Overflow approach for contributors to gain reputation). |
| Proposed Solution | A solution is proposed for a particular problem, but lacks validation. "Must be novel, or at least a significant improvement of an existing technique." |
| Solution with Validation* | Similar to "Proposed Solution" above but includes validation. The proposed solution is evaluated and quantifiable outcomes are reported. |
| Philosophical Paper | Paper that "sketches a new way of looking at things" e.g., provides a conceptual framework or taxonomy. |
| **Research Contribution** (based on Shaw [30] and Paternoster, et al. [23]) | |
| Procedure or Technique | New or better way of doing a particular task. "Not advice or guidelines." Includes methods, approaches or strategies. |
| Qualitative or Descriptive Model | A subjective model such as a structure, taxonomy or framework for a problem area. |
| Empirical Model | Empirical predictive model based on observed data. |
| Tool | Technology, program or application that supports a technique or model. |
| Solution or Answer* | Solution or answer to a problem, which follows a systematic process. This category is left for papers which do not provide a formal model or tool. |
| Report | Interesting observations or rules of thumb, but not sufficiently general or systematic to rise to the level of a descriptive model. |
| * indicates that  modification was made to the original classification scheme (noted above) | |

As noted in Section 3.4, the classification scheme for the topic (to answer RQ2) was created by initially assigning 2-4 relevant keywords that described the topic of the paper. Thereafter, topics were then grouped into categories based on their similarity. This process resulted in eight high-level topic groups (API or Documentation; Community Dynamics and Processes; Community Members; Language; Programming/Development; Question, Answer or Tag; Search; Other). RQ1 was answered using the metrics captured when the studies were compiled (refer to Section 3.3); we examined publications' year and venue to answer this question. As RQ5 required the thorough review and evaluation of each paper, a different classification scheme was developed for this evaluation. This process is explained in Section 3.6.

## 3.6 Quality of Papers

There is growing research interest in the quality of content on Stack Overflow [34]. This is particularly fitting given that this platform is increasingly replacing formal programming languages tutorials and documentation [45]. Multiple studies have thus started to examine Stack Overflow code quality [1, 18, 38], on the premise that the software engineering community could be impacted heavily if code reused from the Stack Overflow platform is low in quality, thus, resulting in the development of defect-prone software. We anticipate that it would be noteworthy to evaluate those studies that are dedicated to this cause to understand their quality profiles in order to support the community's confidence that research conducted on Stack Overflow code are of a high quality. Thus, we created RQ5 (refer to Section 2). In answering RQ2, we observed that 58 papers were dedicated to the *programming/development* topic. These papers were candidates for our quality analysis, an undertaking which demanded the careful review of all of the 58 papers.

We first examined previous protocols (classification schemes) that were used for evaluating research quality, finding that multiple studies were focussed on this cause, with an emphasis on both quantitative and qualitative studies [14, 31]. One such comprehensive classification scheme was developed by Kitchenham and Charters [14], who proposed the evaluation of answers to 59 questions for quantitative studies and 18 for qualitative studies. The questions of the protocol vary in detail for different aspects of research (e.g., design, implementation, analysis and conclusions), with the view of evaluating bias and internal and external validity of research studies.

Further review of the literature revealed that Genc-Nayebi and Abran [7] had previously compressed the criteria developed by Kitchenham and Charters [14] into a new summarised list of categories that were deemed suitable for evaluating research quality (refer to Table 3). To assess the suitability of Genc-Nayebi and Abran [7] protocol for evaluating the quality of the papers under consideration the first two authors randomly selected two papers for evaluation, finding that for some of the dimensions it was easy to assign a *yes* or *no* response (e.g., for dimension A in Table 3), but for others there was need for more scrutiny in coming to a final decision (e.g., for dimension H in Table 3). We thus revisited the earlier guidelines [14, 31], and also examined the textbook recommendations for evaluating research quality [49], in order to extract further details for each question in Table 3. This process eventually led to the delivery of an adaptation of the classification schemes mentioned above, deemed much more exhaustive by the research team. The detailed classification scheme comprising 37 questions which cover multiple facets related to each of the main questions in Table 3 is provided in Appendix B.

To ensure the classification scheme in Appendix B (with compressed version in Table 3) was suitable for analysing the quality of the Stack Overflow papers under consideration, the first two authors conducted another round of pilot testing involving two more randomly selected papers. We again looked to answer each question with *yes* or *no*, which was now easily possible. Thereafter, we discussed our decision for each question, and found full agreement. Of note here is that when using the classification scheme in Appendix B, for those categories with subquestions (e.g., "B" and "C"), the overall category (or dimension) was deemed to be *yes* if answers to all associated subquestions were also answered with *yes*. Being satisfied that this classification scheme was now suitable, we returned all 58 papers to the sample and planned our official reliability assessment. In ensuring reliability of our outcomes, reliability checks were completed by the first two authors checking ten of the 58 papers, against each of the ten quality questions in Table 3, including the consideration of the subquestions in Appendix B. These checks were done in isolation so that the authors were not aware of each other's decision. As above, questions were answered *yes* or *no*, before being compared, with findings revealing that there were seven instances of disagreement out of 100 checks (10 checks each for 10 papers). These instances were then further discussed, which resulted in full agreement.

The remaining 48 papers were then classified by the first author taking the insights of the extensive piloting process and reliability discussions into consideration. The results were subsequently recorded in a Microsoft Excel spreadsheet, with a column allocated for each of the ten quality criteria. For easy statistical analysis these were marked as being met by a particular paper or not, by indicating a '1' (met) or a '0' (not met), slightly different to our "yes" or "no" assessment used during the piloting phase. Results from our paper quality analysis (for answering RQ5) and the other analyses above (for answering RQ1-RQ4) are provided next.

Table 3. Quality evaluation classification scheme for answering RQ5 (based on Genc-Nayebi and Abran [7])

| | | | |
|---|---|---|---|
| A. | Are the aims of the study stated clearly? | F. | Has the diversity of perspectives and contexts been explored? |
| B. | Is the basis of evaluative appraisal clear? | G. | Are there any links between data, interpretation and conclusions? |
| C. | How defensible is the research design? | | |
| D. | Are data collection methods described adequately? | H. | Is the reporting clear and coherent? |
| E. | Has the approach to, and formulation of, analysis been conveyed adequately? | I. | Has the research process been documented adequately? |
| | | J. | Could the study be replicated? |

## 4. RESULTS

This section presents the results of the study, as conducted by the systematic mapping study and quality evaluation, answering RQ1-RQ5. Each subsection is dedicated to answering each of the research questions established (i.e., Section 4.1 provides the results for answering RQ1, Section 4.2 provides the results for answering RQ2, and so on).

### 4.1 Publication Years and Venues (RQ1)

**RQ1**. *(a) What level of academic interest has Stack Overflow generated over time, and (b) what publication venues are targeted?* Of the initial 1,084 papers identified, 265 withstood the rigours of our inclusion/exclusion process and are now used to answer our research questions (i.e., RQ1 to RQ4). Of these 265 papers, the first research papers appeared in a publication venue during 2011 (P51 and P127 in Appendix A). The numbers of publications focussed on Stack Overflow then increased in the next year (in 2012), with at least 40 papers in each of 2013 and 2014, shown in Figure 2(a). Interestingly, this doubled to over 80 papers in both 2015 and 2016, with a slight decrease from 2015 to 2016 (refer to Appendix A for further details). As shown in Figure 2(b), a clear majority of the papers were conference papers (201 papers or 75.9%), with less than 10% published by journals (25 papers altogether). Of the 265 total papers, 20 were published as a part of symposia while 15 papers were published in workshops. The fewest Stack Overflow papers were published as part of a book chapter (only one paper, P33), with three papers not categorised as being published in a scholarly recognised venue (P78, P93 and P196).

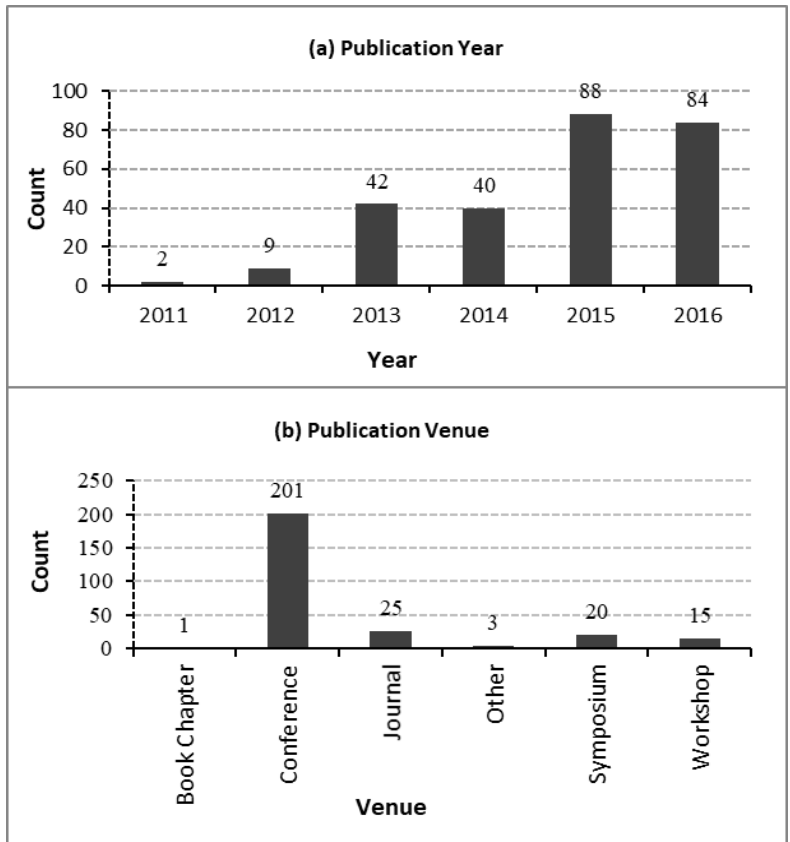

Figure 2. Publication counts by year (a) and venue (b)

Altogether 114 venues were targeted with Stack Overflow publications, with the largest number of papers published as part of the IEEE Working Conference on Mining Software Repositories (40 papers) and IEEE International Conference on Software Engineering (13 papers). In the assessment of journals, Stack Overflow papers were mostly published in Empirical Software Engineering (4 papers) and Journal of Information Science (3 papers). The International Symposium on Foundations of Software Engineering was also popular with Stack Overflow papers (6 papers), and CEUR Workshop Proceedings (4 papers). The full list of venues are available in Appendix C.

## 4.2 Research Topics (RQ2)

**RQ2**. *What software engineering topics are most frequently explored on Stack Overflow?* As noted in the previous section, our extraction of keywords from the papers led to the grouping of keywords into eight categories (see a description of these categories in Table 4). These categories and their frequencies are shown in Figure 3. Here it is shown that Stack Overflow papers analysing *Questions, Answers or Tags* were most frequent (61 papers or 23.0%, e.g., P5, P36 and P229), followed by *Programming/Development* (58 papers or 21.9%, e.g., P86, P153 and P199), *Community Dynamics and Processes* (55 papers or 20.8%, e.g., P43, P152 and P206) and *Community Members* (53 papers or 20.0%, P8, P154 and P157). In fact, there is a clear split between these categories (the top four) and the remaining four (*API or Documentation*, *Language*, *Search* and *Other*) in Figure 3.

In terms of the actual Stack Overflow content that was researched in the categories, Table 4 provides detailed descriptions of the topics. For instance, papers that were focussed on *Question, Answer or Tags* were focussed on the question, answer and tag components of Stack Overflow posts, exploring various issues related to these aspects. *Programming/Development* papers were dedicated to the study of Stack Overflow code aspects, including the evaluation of source code on Stack Overflow. The *Community Dynamics and Processes* topic was comprised of papers about the general Stack Overflow community dynamics, including the aspects around the rules of the community, and *Community Members* studied various properties of the contributors on Stack Overflow, under the human aspects theme. Considering the other topics, *API/Documentation* (e.g., P11, P50 and P110), *Language* (e.g., P40, P188 and P249) and *Search* (e.g., P80, P120 and P128) are all self-explanatory in Table 4, albeit these papers featured less prominently in Figure 3. In addition, only 7 of the 265 papers (2.6%) did not fit into a category.

**Table 4. Summary topics and descriptions**

| Topic | Description |
|---|---|
| API or Documentation | Papers that investigate aspects relating to API or Documentation such as API changes or deprecation, deficient documentation and effects of particular types of documentation. |
| Community Dynamics and Processes | Papers that look at overall community dynamics which can include general CQA processes or specific community dynamics on Stack Overflow. |
| Community Members | Papers that investigate the characteristics of Stack Overflow community members including aspects such as user reputation, user learning and the search for experts. |
| Language Use | Papers that investigate aspects relating to the use of Software Engineering language. For example, retrieving questions from other languages or sentiment detection. |
| Programming/Development | Papers that relate to programming or development, including coding and systems development. Papers in this category include evaluating source code and understanding project requirements. |
| Questions, Answers or Tags | Papers that look at the question, answer or tag elements of the Stack Overflow website. This category include papers covering the characteristics of one or more of these elements, such as Quality. |
| Search | Papers that investigate search and/or querying to improve the query or the appropriateness of the results from a query executed on Stack Overflow. |
| Other | Other papers that could not fit into the categories above. This included papers such as Stack Overflow Developer Myths [68]. |

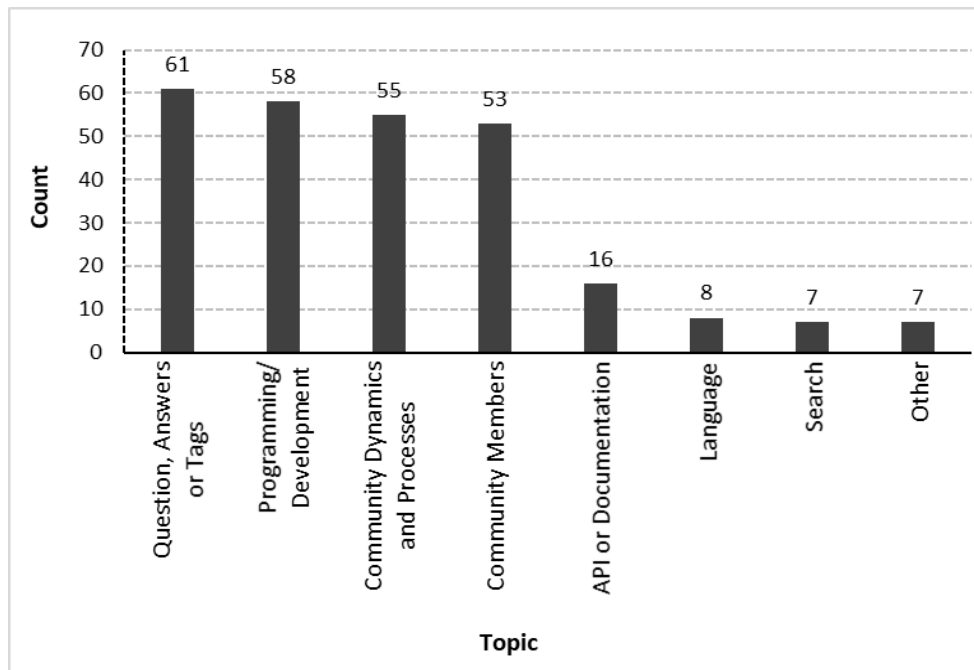

**Figure 3. Count of topics**

### 4.3 Research Approaches (RQ3)

**RQ3**. *What research approach(es) are used to investigate these topics?* The research approaches used to investigate various topics are shown in Figure 4. The x-axis shows the research approaches employed (e.g., Proposed Solution) and the y-axis shows the topics (e.g., Programming/Development). The size of circles (or bubbles) that are at the intersection of certain x- and y-axes values represent the number of papers that satisfy the corresponding criteria depicted in these axes. Interestingly, Figure 4 shows that nearly half (45.3%) of the published papers were *Evaluation* type papers (120 papers altogether, e.g., P20, P211 and P240). In particular, *Evaluation* papers that considered *Community Dynamics and Processes* made up 15.1% of all the papers published across venues in Figure 2(b). We noted that *Evaluation* papers for *Programming/Development* made up 9.4% of all papers in Figure 4. As Stack Overflow is relatively new, this outcome is unsurprising. While 87 papers (32.8%) were *Proposed Solution* (e.g., P46, P54 and P218), an additional 46 papers (17.4%) proposed a solution with some form of validation in Figure 4 (see *Solution with Validation*, e.g., P233, P236 and P258). This means that only 34.6% (46 out of 133) of the published work that proposed a solution using Stack Overflow data provided some form of quantifiable validation. *Philosophical* papers made up a small proportion of the papers (4.5%, e.g., P136, P180 and P247), where these papers often examined *Community Dynamics and Processes* or *Community Members*, or studied some aspects related to Stack Overflow *Programming/Development* or *Question, Answer or Tags*.

### 4.4 Contribution Type (RQ4)

**RQ4.** *What form did the contribution of the research take?* Six types of research contributions were classified, as shown in Figure 5. Unlike *research approach*, the types of contributions that were provided by researchers studying content on the Stack Overflow platform varied. A majority (75 papers or 28.3%) of papers gave a *Solution or Answer* (to a question) as their contribution (e.g., P73, P152 and P158). This was followed by *Procedure or Technique* (57 papers or 21.5%, e.g., P96, P141 and P195) and *Tool* (50 papers or 18.9%, e.g., P149, P170 and P184). Models altogether made up 22.3% (59 papers) of the total paper cohort, with 43 of these being *Empirical Model* (e.g., P29, P55 and P84). When examining the topics, one again, *Community Dynamics and Processes* made up the largest individual grouping of 8.7% (23 in Figure 5) of the total papers when considered in relation to the *Solution or Answer* papers contributed by the Stack Overflow community. This was closely followed by papers in the *Programming/Development* category for *Tool* contributions (20 papers). Papers focussed on the *Questions, Answers or Tags* topic that contributed *Empirical Models* recorded a similar number (19 papers, representing 7.2%). In Figure 5 we see a good spread the contributions.

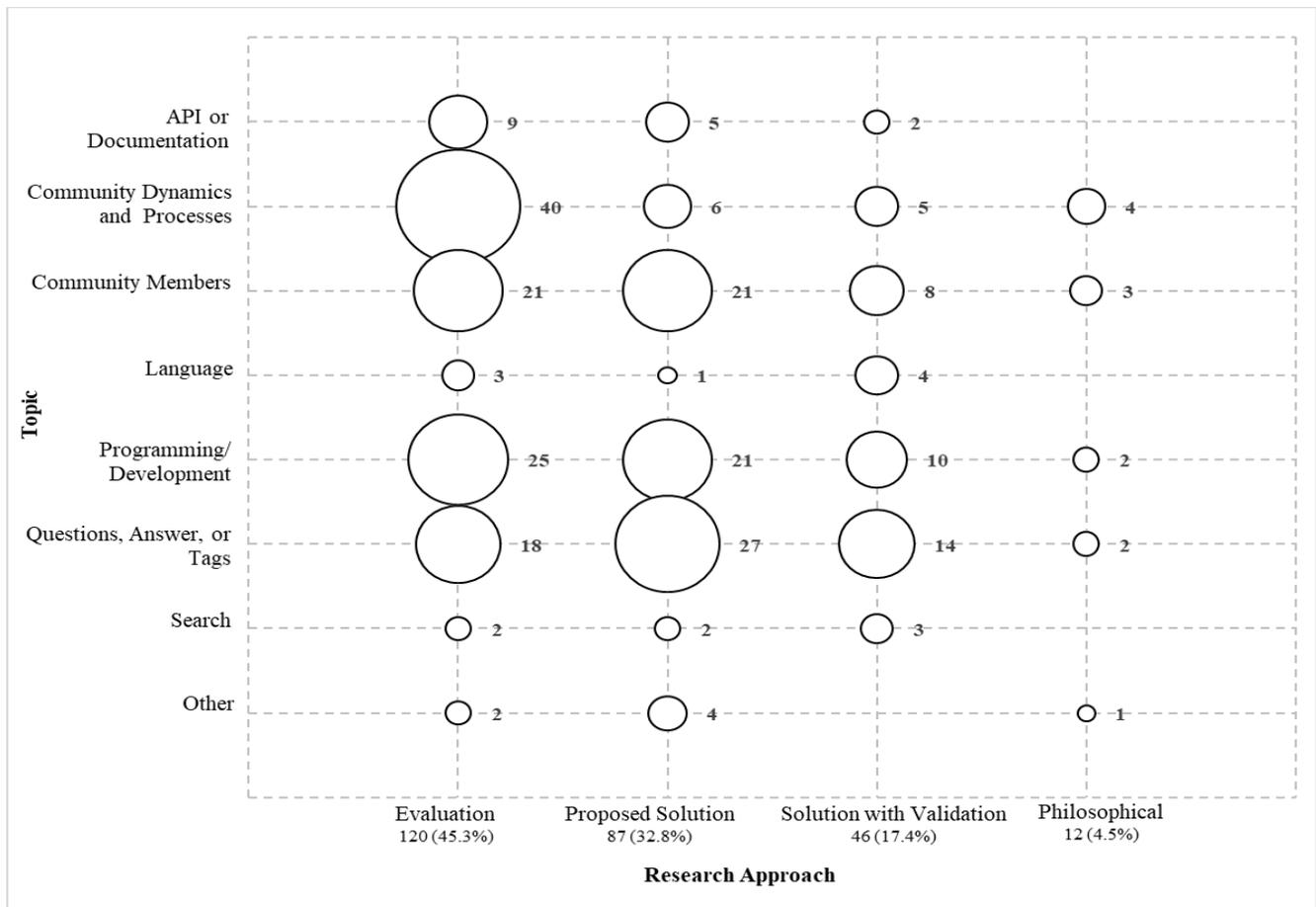

**Figure 4. Systematic map - topic and research approach**

### 4.5 Code Paper Quality (RQ5)

**RQ5.** *What is the quality of research conducted on Stack Overflow?* In thoroughly reviewing 58 papers under the *Programming/Development* topic we observed that 33 (57%) of these Stack Overflow papers related generally to the code provided on Stack Overflow (e.g., P153, P179 and P232), while 8 papers considered bugs or errors in code (e.g., P44, P105 and P250). The remaining 17 papers were dedicated to specific categories of programming resources provided on Stack Overflow (e.g., P13, P34 and P85). For example, one paper (P199) examined programming issues faced by Android software development practitioners. A majority of the 58 papers were published in 2015, with 19 papers in total published in this year. This was followed by 16 Stack Overflow *Programming/Development* papers in 2016. From 2012 to 2014 there were 3 papers, 11 papers, and 9 papers, respectively. Here we see a good spread of studies focussed on Stack Overflow coding resources over time.

We performed detailed evaluations of the papers' quality using the classification scheme in Appendix B (see summary also in Table 3). The outcomes of these evaluations are provided in Table 5. On average the papers met 5.8 out of the ten quality aspects which were considered, with a standard deviation of 2.7 and median of 6.0. Unsurprisingly, quality criterion A, '*Are the aims of the study stated clearly?*', was met the most of all the quality dimensions with 84.5%, followed by H '*Is the reporting clear and coherent?*' with 77.6% of the papers evaluated meeting this criterion. Quality criterion F, '*Has the diversity of perspectives and contexts been explored?*', was met the least (just 36.2% of papers met this criterion), closely followed by I, '*Has the research process been documented adequately?*' (with 38.7% of papers satisfying this dimension). The dimension B, '*Is the basis of evaluative appraisal clear?*', was slightly higher than I, with 39.7% of papers satisfying this criterion (refer to Table 5).

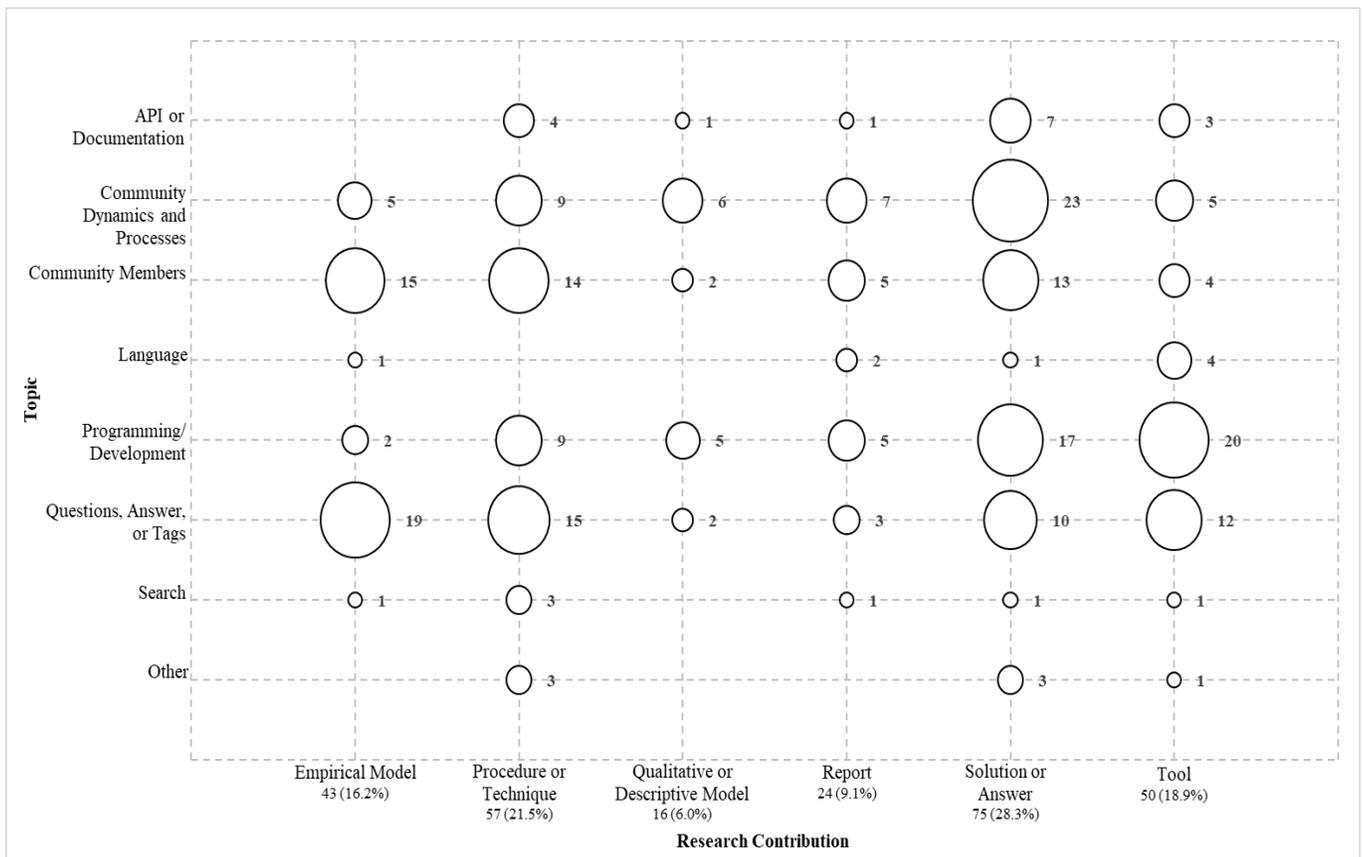

**Figure 5. Systematic map - topic and contribution**

A cursory check of our quality evaluation outcomes suggested that the quality of the research papers that were published on Stack Overflow related to the length of the paper. We thus formally examined this evidence in Table 6. Here it is shown that while the trend of longer Stack Overflow papers having higher quality is somewhat evident, there are some exceptions. For instance, in Table 6 the papers having length 13 (P56 and P78) and 18 pages (P116) had an average quality of 5. Formal Spearman correlation tests [48] were conducted to explore the quality of the papers more widely, as our distributions violated normality assumption [47]. Table 7 shows correlation outcomes for the various quality dimensions, when papers were published, and the length of the papers. We also correlated these aspects against the assessment of Stack Overflow papers' overall quality. Results here reveal that longer Stack Overflow papers were of higher quality overall ($r= 0.65$, and statistically significant). This pattern of outcome also held for seven of the 10 quality dimensions in Table 7, with particularly high correlation for dimensions G (*Are there any links between data, interpretation and conclusions?*) and I (*Has the research process been documented adequately?*); $r=0.66$ and $r=0.63$ respectively. Multiple other quality dimensions were particularly influential in determining Stack Overflow papers' overall quality, with quality criterion I having the highest relationship with the overall quality score ($r=0.77$). Other notable outcomes were seen for criteria C, *'How defensible is the research design?'* ($r=0.70$), and J, *'Could the study be replicated?'* ($r=0.73$). We did not observe the year of publication to be related to the overall quality of papers, or any individual dimensions.

**Table 5. Quality of papers**

| Quality Criterion[6] | Number of Papers | Percentage of Papers |
|---|---|---|
| A. Are the aims of the study stated clearly? | 49 | 84.5% |
| B. Is the basis of evaluative appraisal clear? | 23 | 39.7% |
| C. How defensible is the research design? | 29 | 50.0% |
| D. Are data collection methods described adequately? | 34 | 58.6% |
| E. Has the approach to, and formulation of, analysis been conveyed adequately? | 41 | 70.7% |
| F. Has the diversity of perspectives and contexts been explored? | 21 | 36.2% |
| G. Are there any links between data, interpretation and conclusions? | 28 | 48.3% |
| H. Is the reporting clear and coherent? | 45 | 77.6% |
| I. Has the research process been documented adequately? | 23 | 39.7% |
| J. Could the study be replicated? | 41 | 70.7% |
| *Average* | *33.4* | *57.6%* |

**Table 6. Average paper quality and page length**

| Number of Pages (length) | Number of Papers | Average Quality |
|---|---|---|
| 2 | 1 | 2 |
| 4 | 15 | 3.6 |
| 5 | 7 | 4 |
| 6 | 4 | 5 |
| 7 | 1 | 5 |
| 8 | 1 | 9 |
| 9 | 1 | 7 |
| 10 | 14 | 7.2 |
| 11 | 2 | 8 |
| 12 | 3 | 9.3 |
| 13 | 2 | 5 |
| 14 | 1 | 10 |
| 15 | 2 | 7 |
| 18 | 1 | 5 |
| 20 | 1 | 7 |
| 32 | 1 | 8 |
| 42 | 1 | 10 |

We next examine the quality of the *Programming/Development* papers in relation to the research approaches that were used and the contributions that were provided (i.e., the outcomes above for RQ3 and RQ4). Table 8 shows the average quality of the papers for each research approach that was adopted. While a large number of papers were either *Evaluation* or *Proposed Solution* papers (25 and 21 papers respectively), there is a notable difference between the quality of *Proposed Solution* papers and papers that proposed a *Solution with Validation.* In Table 8 it is noted that the latter Stack Overflow papers were of a much higher quality; 7.3 versus 4.4 (an average three points more). This outcome is somewhat expected, considering that papers that were categorised as *Solution with Validation* actually evaluated a solution that was developed, and thus, signalled if the outcomes provided were of actual utility. We inspect the outcome for the average quality of the *Programming/Development* papers in relation to the contributions that were provided in Table 9. Here it is noticed that papers that contributed *Tool*, *Report* and *Empirical Model* had below the average quality (4.9, 4.2 and 4.5 respectively, as against the average of 5.8 mentioned above). In contrast, outcomes in Table 9 show that papers that contributed a *Procedure or Technique*

---

[6] Refer to Appendix B for full list of questions.

(with an average quality score of 7), *Qualitative Descriptive Model* (with an average quality score of 6.4) and *Solution or Answer* (with an average quality score of 6.5) had above the average quality. Studies in our sample did not look at providing mechanisms to maintain or repair code resources on Stack Overflow. The full dataset including categories and quality information for each paper are provided in Appendix D. We discuss these outcomes and those above in the following section.

**Table 7. Paper quality correlation table**

|  | Year | length | A | B | C | D | E | F | G | H | I | J | Quality |
|---|---|---|---|---|---|---|---|---|---|---|---|---|---|
| **Year** | 1 | 0.25 | -0.14 | 0.23 | 0.17 | -0.05 | 0.13 | -0.13 | 0.13 | 0.02 | 0.24 | 0.02 | 0.13 |
| **length** |  | 1 | 0.19 | *0.28 | *0.47 | 0.18 | *0.35 | *0.26 | **0.66** | 0.16 | **0.63** | *0.46 | **0.65** |
| **A** |  |  | 1 | -0.04 | *0.33 | 0.12 | *0.35 | 0.22 | 0.13 | 0.23 | 0.25 | *0.46 | *0.49 |
| **B** |  |  |  | 1 | 0.18 | 0.04 | -0.10 | 0.05 | 0.20 | 0.10 | *0.35 | 0.21 | *0.37 |
| **C** |  |  |  |  | 1 | 0.14 | *0.42 | *0.32 | *0.35 | *0.29 | *0.46 | *0.49 | **0.70** |
| **D** |  |  |  |  |  | 1 | 0.15 | 0.20 | 0.18 | 0.05 | *0.32 | *0.38 | *0.45 |
| **E** |  |  |  |  |  |  | 1 | 0.25 | *0.32 | 0.02 | *0.44 | *0.42 | **0.57** |
| **F** |  |  |  |  |  |  |  | 1 | *0.28 | *0.32 | *0.27 | *0.33 | **0.57** |
| **G** |  |  |  |  |  |  |  |  | 1 | 0.11 | **0.56** | *0.32 | **0.63** |
| **H** |  |  |  |  |  |  |  |  |  | 1 | 0.10 | 0.20 | *0.39 |
| **I** |  |  |  |  |  |  |  |  |  |  | 1 | **0.52** | **0.77** |
| **J** |  |  |  |  |  |  |  |  |  |  |  | 1 | **0.73** |
| **Quality** |  |  |  |  |  |  |  |  |  |  |  |  | 1 |

* indicates significance ($p < 0.05$), *Italics* indicates medium effect size using Cohen's classification ($0.3 \leq r \leq 0.49$), **bold** indicates high effect size using Cohen's classification ($r \geq 0.50$) [58]

The questions corresponding to the quality dimensions A to J that appear in column headers are provided below, with full classification scheme in Appendix B.

A. Are the aims of the study stated clearly?
B. Is the basis of evaluative appraisal clear?
C. How defensible is the research design?
D. Are data collection methods described adequately?
E. Has the approach to, and formulation of, analysis been conveyed adequately?
F. Has the diversity of perspectives and contexts been explored?
G. Are there any links between data, interpretation and conclusions?
H. Is the reporting clear and coherent?
I. Has the research process been documented adequately?
J. Could the study be replicated?

**Table 8. Average paper quality for research approach**

| Research Approaches | Number of Papers | Average Quality |
|---|---|---|
| Evaluation | 25 | 6.2 |
| Philosophical | 2 | 6.5 |
| Proposed Solution | 21 | 4.4 |
| Solution with Validation | 10 | 7.3 |

Table 9. Average paper quality for contribution type

| Contribution Type | Number of Papers | Average Quality |
|---|---|---|
| Empirical Model | 2 | 4.5 |
| Procedure or Technique | 9 | 7 |
| Qualitative or Descriptive model | 5 | 6.4 |
| Report | 5 | 4.2 |
| Solution or answer | 17 | 6.5 |
| Tool | 20 | 4.9 |

## 5. DISCUSSION AND IMPLICTIONS

In harnessing the power of the crowd, online platforms such as wikis (e.g., Wikipedia) and forums or community question answering (CQA) portals are becoming widespread [12]. CQAs in particular help users to solve very difficult problems, that may be of a personal or professional nature [39]. Anecdotal evidence suggests that software engineers, and the software development community in general, depend on Stack Overflow for their domain specific solutions [9], and this platform has also been the subject of academic research [19]. Our work here aimed at gaining an understanding of the academic interest that is dedicated to the study of Stack Overflow, and it is fitting to observe that the academic community has researched various aspects of this community portal. Research focussed on Stack Overflow covers aspects related to both the resources that are provided and those that provide such resources and the processes they use. The fact that previous work has found that 84.4% of software developers' Google searches resulted in a suitable Stack Overflow answer on the first page of the results pool that is returned [22] demands that such answers are reliable, as the opposite could be catastrophic for the software engineering community. Researchers have noted that issues related to gamification and information quality is often a cause for concern [5]. For instance, suggestions/recommendations for a specific API use which has critical security flaws could quickly propagate across the software engineering community, resulting in the development of insecure software that could subsequently result in major costs to businesses that employ such solutions. Research is thus needed to understand the scale and nature of academic work that is being performed to ensure that the research is properly conducted and can assure that information provided in answers (and questions) is relevant, up-to-date and of high quality. Research questions crafted in this work are aimed at initiating this process, in checking to see what research has studied Stack Overflow.

In this study, we attempt to answer five research questions (RQ1-RQ5), aimed at confirming the popularity of Stack Overflow and surveying the range of studies that are conducted by academics, including: the publication year and venue, research topics, research types and research contributions. We then evaluate the quality of all studies dedicated to Stack Overflow programming resources. It should be noted that there has been prior work in the general area of CQA systems such as the work of Srba and Belikova [32], however; our work here focuses on the popular Stack Overflow platform. Also, none of the prior work have evaluated the quality of research conducted on this platform which offers insights into future directions. We now consider the outcomes for each of these enquiries in turn in the following five subsections.

### 5.1 Publication Years and Venues (RQ1)

**RQ1**. *(a) What level of academic interest has Stack Overflow generated over time, and (b) what publication venues are targeted?* Our study shows that interest in Stack Overflow has increased since the first published papers in 2011. In particular, there were two significant increases in publications dedicated to the Stack Overflow platform between the years 2012 and 2013 (where research increased from 9 to 42) and from 2014 to 2015 where research doubled. Our outcomes show that the popular venue or publication type for these papers is conferences with 201 out of 265 papers published in conferences. Journals, Symposiums and workshops each accounted for less than 10% of the papers that studied Stack Overflow. This somewhat aligns with Srba and Bielikova [32] outcomes, which found an increase in CQA research articles from 2005 to 2014, however; these authors found a steadier increase during this time. Additionally, these authors noted that a significant number of papers in their study came from conferences, and that a small number of papers in the study were from journals. The fact that there is such high level of academic interest given to Stack Overflow is noteworthy, and is perhaps linked to the accessibility and availability of the platform's data.

This accessibility and availability of data enables continuous research of Stack Overflow, which is important to maintain the long-term sustainability of Stack Overflow and other CQA platforms. This is important because adapting to the changes of the community and the needs of the members is likely to ensure that there is continued interest in the platform, and users are likely to continue to return to these websites. This is particularly significant as a larger number of users help to ensure websites such as Stack Overflow runs smoothly, and particularly in terms of the provision of a large number of good quality questions and answers. The increased research interest in Stack Overflow highlights that software engineering researchers have a strong curiosity in studying and improving outcomes on the community-driven Stack Overflow platform. In this regard, we consider the nature of topics that were considered by studies that have researched Stack Overflow in the following section.

### 5.2 Research Topics (RQ2)

**RQ2**. *What software engineering topics are most frequently explored on Stack Overflow?* Our findings show that most studies focussed on *Question, Answer* or *Tags,* making these the most common Stack Overflow topics that are investigated (refer to Figure 3). These topics were followed by *Programming/Development*, *Community Dynamics and Processes*, and *Community Members*. By and large the aforementioned topics dominated research interest on Stack Overflow. The remaining topics received little attention, including *API or Documentation*, *Language*, *Search* and *Other*, which had only 16, 8, 7 and 7 papers respectively.

Interestingly, the top four topics are directly related to CQA. Community ('C' in CQA) reflect *Community Dynamics and Processes* and *Community Members*. As a successful example of CQA, Stack Overflow has not only been researched to learn more about programming and software engineering, but to learn about CQA in general for improving other CQA sites. In fact, just as the human factors in software engineering has been gaining significant attention [16, 17], we see a similar pattern in the level of interest placed on studying these aspects of Stack Overflow (Community Dynamics=55 studies and Community Members=53 studies, or 108 papers focusing on aspects related to human factors and processes altogether). Gamification, for example, uses elements of "game design in non-game environments to induce certain behaviour" [24] such as improving user participation, a technique that Stack Overflow uses and other sites can learn from. Question Answering ('QA' in CQA) is directly represented by the topic *Question, Answer and Tags*. Additionally, *the topic of the 'QA' is directly related to the second most popular topic of papers - programming/development. Question, Answer and Tag* papers often focus on improving CQA aspects, or making the Stack Overflow software engineering knowledge base more precise and relevant. This could be in the form of improving the quality of these aspects or the process, in making Stack Overflow easier to use (e.g., how to use keywords for tagging to enable easy and precise searches for members' queries). The need for quality has been emphasized in the previous subsection. In particular, the quality of answers (in response to questions) is significant to the software engineering community in that this points to the source of information users are seeking, and are likely to reuse [46]. Additionally, the quality of questions and the language used when creating tags can affect the process and potential information that is provided (or not provided) by other users [26, 35].

Interestingly, while *API* or *Documentation* was a topic of 16 of the research papers mapped in this work, it is interesting to note that Stack Overflow has developed Stack Overflow Documentation[7]. Released as a beta version in mid-2016, Stack Overflow Documentation is an extension of the regular Stack Overflow platform that, unlike most CQA questions which are specific to a particular problem, focuses on broad understanding of topics. While this work did not focus specifically on this aspect of Stack Overflow, it would be interesting for research work to explore the usefulness of the Stack Overflow Documentation platform to the software engineering community. That said, of interest in the work was the main Stack Overflow platform. As a platform of reference for other CQA websites, understanding the topics on Stack Overflow reflects the importance of validating Stack Overflow as a platform. However, the online software engineering community is specific, such that lessons learned from Stack Overflow may not always be applicable to other CQA sites. For example, Stack Overflow discourages opinion based questions as these tend not to invoke a definitive 'answer'. To this end, these types of questions are often closed by experienced users, in contrast to Yahoo!Answers which allows such questions. Therefore, it is pertinent that research explores understandings for what types of academic work has (or hasn't) been conducted for the Stack Overflow platform.

---

[7] http://stackoverflow.com/tour/documentation

*Language use* within the community hasn't received much attention among researchers (8 papers). This is a topic that deserves further attention. For example, in the last few years, Stack Overflow has been noted to be a non-welcoming place for new comers due to elitist attitude among existing members. The administrators of the site have openly declared that it is time for change[8]. Particularly, the comments made by some users have been found to be hostile and offensive, and the moderators of the site find it challenging to identify and remove such comments quickly as it requires timely and concerted effort involving the community (i.e., community member flags an offensive comment which is then actioned by the moderator). In 2020, the administrators have issued an open call for researchers to detect abusive comments automatically (e.g., using NLP, sentiment-analysis and machine-learning approaches), by making a dataset of such comments available[9] with the aim of preventing them at their source (i.e., as they are submitted). This may be a fertile avenue for future work. When developed, such automated mechanisms can be of use beyond just the Stack Overflow community (e.g., other Stack Exchange communities and CQA platforms).

Also, research on *searching* relevant content within Stack Overflow (questions, code snippets) has received limited attention (7 papers). Prior work in this area of research have also been limited to the use of traditional machine-learning algorithms (e.g., Naïve Bayes - see P182 in Appendix A). With the development of deep neural-network based algorithms such as ELMo [69] and Google's BERT [70] that understand natural language using contextualised word embeddings that provide better results than existing algorithms, there might be opportunities for leveraging, adapting and extending these in the Stack Overflow domain. While BERT in particular may be able to deliver better results for natural language-based search queries in the Stack Overflow domain (as shown by work in other contexts such as passage re-ranking [71]), it would be worthwhile to evaluate how this might be able to handle queries that contain code snippets. Software engineering researchers have argued that software code is more "natural" (i.e., more repetitive and predictable) than human languages [72]. From that viewpoint, it is possible that BERT's application may provide superior results in this domain. However, this needs to be investigated further. Such developments, if successful, may fast-track other related research activities such as checking for code plagiarism (across different CQA sites), and detecting origins of buggy-code within CQA portals, and also more generally in open source projects.

We next consider the research approaches employed by research work on Stack Overflow and their contributions.

### 5.3 Research Approaches (RQ3)

**RQ3**. *What research approach(es) are used to investigate these topics?* Evidence in this work shows that *Evaluation* papers made up a clear majority of research approaches for studies explored (nearly half, 120 papers), followed by *Proposed Solution* and *Solution with Validation* papers. Philosophical papers made up only 4.5% (12) papers. In fact, papers that proposed a solution made up 50.2% of papers (133 out of 265), with 34.6% (46 out of 133) of these including a validation (*Solution with Validation*). This means that approximately a third of the solutions provided by academic work that has focussed on Stack Overflow contained quantifiable validation. This outcome suggests that there is scope to perform validation on the solutions (techniques provided by 87 studies that lacked validation) that are provided by academics for Stack Overflow. This would be particularly necessary if such solutions are in response to the challenges introduced by Stack Overflow community policies and norms (e.g., community voting norms), where the drive is to enhance overall outcomes provided for the software engineering community.

In fact, Srba and Bielikova [33] found that the quality of answers on Stack Overflow was decreasing, with churn rates increasing, supporting the importance of evaluation papers to understand Stack Overflow and how the platform and its community change over time. In addition, the lack of a high level of validated solutions noted above to solve potential problems on Stack Overflow (noted by Srba and Bielikova [33]) is of concern. Over time, however, the proportion of evaluation papers has been decreasing. Our evidence shows that both papers published on Stack Overflow in 2011 were evaluative (i.e., 100%), while over 66% of this type of articles were published in 2012.

---

[8] https://stackoverflow.blog/2018/04/26/stack-overflow-isnt-very-welcoming-its-time-for-that-to-change/

[9] https://www.workshopononlineabuse.com/resources-and-policies/resources

Thereafter, the number of this form of Stack Overflow publications dropped each subsequent year (58% in 2013, 48% in 2014, 40% in 2015), with 39% published in 2016. On the other hand, a more positive trend noted in our study is that the number of *Solution with Validation* type papers are slowly increasing, especially when evaluated in relation to the proportion of papers overall. While our evidence suggests that there were no such papers in 2011 or 2012 that provided a *solution with validation*, in the years following the trend for this form of article continued to increase slowly, with 12% in 2013, 18% in 2014, 19% in 2015 and 23% in 2016. These findings suggest that we may see improvements in the nature of work that is dedicated to validating the Stack Overflow platform, and particularly if this trend continues.

Our findings are in fact supported by Shah, et al. [29], who identified evaluation as one of three challenges for research on social question and answer forums. Given that Stack Overflow was created in 2008, it makes sense that initially papers would seek to evaluate the platform, and then consider improvements by proposing solutions. Additionally, it is unsurprising that philosophical papers represented a small number of papers, as understanding is needed before it is possible to create 'new ways of looking a things' (i.e., philosophical reflections). However, with many solutions being proposed already for the Stack Overflow community, there is a need to understand how these improve on the status quo, or the improvement opportunity they offer. Further investigations into the quality and validity of solutions that are provided are necessary, so that these solutions can be employed by Stack Overflow or similar CQA websites. In line with this viewpoint, we next evaluate the contributions of the studies researching Stack Overflow to identify the types of outputs that are provided to the community by the research studies reviewed.

## 5.4 Contribution Type (RQ4)

*RQ4. What form did the contribution of the research take?* Our outcomes revealed that *Solution or Answer* and *Procedure or Technique* papers were most common contribution types for research focusing on Stack Overflow. These contributions were followed by *Tools* and *Empirical Models*. Only a small number of papers contributed *Reports* or *Qualitative or Descriptive Models*. Aligned with the discussion for RQ4, it is unsurprising to find *Solution or Answer* as the most common contribution. In fact, 70 out of 75 total *Solution or Answer* papers were *Evaluation* type studies. Similarly, in examining the trends further we noted that the proportion of *Solution or Answer* papers is decreasing overall (2011=0%, 2012=56%, 2013=35%, 2014=37%, 2015=21%, 2016=26%), while the proportion of *Procedure or Technique* papers is increasing overall (2011=0%, 2012=11%, 2013=9%, 2014=14%, 2015=27%, 2016=26%).

Research agenda questions suggested by Shah, et al. [29] should focus on the understanding of question and answer forums and research outcomes dedicated to these platforms, as it was noted that this is a new area of research. These authors suggested questions such as: 'What causes people to be involved in social Q&A?, 'What are the factors that need to be considered when measuring the quality of questions and answers?' and 'Why some services fail, why others succeed?'. These questions would help us to understand the motivations of users, the quality of information provided by contributors to Q&A sites, and the differences between Q&A platforms and the utility they provide. These types of questions suggest that *Solution or Answer* type contributions are needed, before *Tools* or *Models*. For example, in keeping with the view that an understanding of a given research area is required before *Tools* or *Models* are created. *Tools* in particular are only useful after a problem is fully understood and opportunities evaluated. That said, *Models* may help us to understand the state of affairs and trends and patterns that could inform tool design. Having said that, there is need for contributions which help to improve Stack Overflow and CQA platforms, such as tools and techniques. These contributions are necessary to the success of Stack Overflow, which relies on adapting to changing community behaviours [19]. Given these findings, the quality of these contributions is evaluated by assessing the quality of Stack Overflow's *Programming/Development* papers.

## 5.5 Code Paper Quality (RQ5)

*RQ5. What is the quality of research conducted on Stack Overflow?* On average the 58 *Programming/Development* papers that were assessed for quality met 5.8 of the ten quality criteria outlined by Genc-Nayebi and Abran [7], with the median paper quality being six out of ten. Our findings here are worrying when considering that software developers rely heavily on the Stack Overflow platform [40], and the research community is often tasked with helping to validate the utility and reliability of Stack Overflow. The fact that some studies are not properly designed to meet with research quality guideline is troubling, and particularly in terms of the potential to accept recommendations of researchers for aiding software engineering practice.

In fact, in the evaluation of *Programming/Development* type research, quality criterion A, *Are the aims of the study stated clearly?*, was met the most, whilst quality criterion F, *Has the diversity of perspectives and contexts been explored?*, was met the least. We also observed that there is a relationship between paper quality and length of the paper, as indicated with a correlation of $r=0.65$, although 45.8% of papers contained six or less pages. This outcome is not surprising, as it could be expected that papers that are shorter in length would be unable to meet as many of the quality questions, and such studies are quite often published as work in progress and new and emerging research. However, there was one paper that was 18 pages in length that only met five (or half) of the research quality criteria, and a paper with 13 pages in length that had the lowest quality (equal with two papers which had lengths of four and five pages), meeting only one of the ten quality questions. It should be expected that such longer papers would meet a majority of the quality questions. As the average paper length in this study was 9 pages, the average quality of 57.6% is disappointing. While some shorter papers are able to meet a large number of criteria, there were four longer papers (ten pages or more) which fail to meet more than half of the criteria that are applicable to properly conducted research studies. In considering this outcome, it should also be noted that 20 of the 58 papers contributed a tool in their work. These papers often failed to meet the given quality criteria as they often followed a different paper structure to that recommended by research guidelines [7, 14, 49]. This suggests that there is inconsistency among the authors of papers, and that care should be taken by the research community in selecting appropriate papers for references and for which theoretical support and research design assumptions are made.

Additional to these outcomes and syntheses, the quality of papers was compared to the research and contribution types. When considering the contribution types of the papers, papers which contributed a *procedure or technique* had the highest average quality. Importantly, papers that propose a *solution with validation* have higher quality than those without. In fact, papers that propose a *solution with validation* met on average three more quality questions. This suggests that researchers should seek to provide validation for their Stack Overflow solutions in the delivery of higher quality work. Our study also observes gaps in Stack Overflow research in terms of mechanisms to help with enhancing the quality of data on the platform, and especially code quality. Even though a range of programming topics are studied, these largely study code snippets and the content of the question and answer text that is provided. Limited effort was observed that look at improving the quality of content on the Stack Overflow platform itself. This gap provides an opportunity for future work.

Considering these findings in relation to Meldrum, et al.'s [20] preliminary findings, in terms of insights for the conduct of systematic reviews and mapping studies, outcomes here suggest that the categorisation of papers using the abstract for studying research types, as is commonly done in the conduct of systematic mapping studies [52], may be insightful for indicating the quality of research papers as a whole. This proposition is forwarded because we observed that abstracts that led to the classification of a paper as a *solution with validation* had a higher quality than those classified as a *proposed solution*. While intuitively an abstract may be used to decide on the quality of a paper, the alignment of the research approach and paper quality results confirms the importance of an abstract as an indicator of a paper's worth as a whole.

## 6. THREATS

Our study followed the systematic mapping process as described by Peterson, et al. [25], and where necessary, when we could not satisfactorily extract appropriate content from abstracts we have reviewed papers following systematic review guidelines [14]. As with studies using these approaches, there are some threats to validity that should be considered when evaluating the outcomes that are provided in this work [25, 52].

The selection of primary papers can be seen as a threat, as we could have missed relevant papers. That said, to mitigate the likelihood of this threat materialising, searches were conducted on six digital libraries, and we used broad search strings (e.g., 'Stack Overflow' and 'StackOverflow'). Proceedings from these search results were also checked for additional relevant papers, which only resulted in one additional *new* paper. This meant that we were likely to have surveyed an exhaustive list of studies, rather than miss pertinent ones. This, in turn, presented a paper selection threat. Therefore, formal reliability checks were conducted to ensure agreement on the inclusion and exclusion of papers. Additionally, pilot evaluations were conducted to ensure that all authors had the same understanding, and thus, the likelihood of a very reliable mapping study and quality evaluation.

Classifying the papers by using the abstract can be seen as a threat to reliability, even though it is a common practice when creating a systematic map. Peterson, et al. [25] recommended "adaptive reading depth for classification" as abstracts can miss certain pieces of information. In our study, when there was ambiguity in the abstracts we reviewed the paper in full. This meant that our understanding of the abstract could be checked against the actual contents of the paper, thus, reducing the threat to reliability. We have also read all of the *Programming/Development* papers (58 in total) and critically evaluated these. Furthermore, research involving human judgement is subjective, and so questions naturally arise regarding the validity and reliability of the outcomes of our classifications. To mitigate this threat, we employed systematic methods to derive our topics, research approaches, research contributions and paper quality evaluations, which were assessed for both accuracy and objectivity via reliability checks (as reported in Section 3.6).

Finally, construct validity aims to verify that tests used to measure a phenomenon actually does so [36]. To mitigate construct validity as a possible threat, we have adapted numerous criteria. In answering RQ3 and RQ4 we have adapted guidelines of Wieringa, et al. [37], Shaw [30] and Paternoster, et al. [23], which were previously assessed as suitable for classifying research approaches and contributions [55, 56, 57]. In answering RQ5 we have adapted and expanded the compressed questions provided by Genc-Nayebi and Abran [7], which were previously chosen as a subset of those provided by Kitchenham and Charters [14]. These criteria were previously used and assessed for accuracy as part of guidelines for conducting systematic reviews and evaluating research quality [2, 10].

## 7. CONCLUSION AND FUTURE RESEARCH

In this paper we conducted a systematic mapping study and quality evaluation on the crowdsourced knowledge platform Stack Overflow to understand and validate the research that has been conducted on this platform thus far. Software engineering has a particular reliance on CQA platforms, and particularly with the help that is offered by programming code included in solutions (or answers) provided by contributors. With significant interest in CQAs, and Stack Overflow in particular, it is necessary for academic work to ensure that information provided in answers (and questions) on Stack Overflow is relevant, up-to-date and of high quality. Ongoing research interest may indeed address this gap; however, previous work did not review the scale of scientific attention that is given to this cause. We have looked to do so in this systematic mapping study and quality evaluation, exploring the level of academic interest that is given to Stack Overflow, the topics that are researched, the approaches that are used, the contributions that are provided by research studies, and the quality of *Programming/Development* papers.

Our outcomes show that interest in Stack Overflow has increased since the first published papers in 2011, and particularly in terms of the number of conference papers that are published. We observed that most studies focused on Questions, Answers or Tags, making these the most common Stack Overflow topics that are investigated. However, there was also strong interest in Stack Overflow's Programming/Development, Community Dynamics and Processes, and Community Members. Understudied areas include API/documentation, language used and search. Opportunities for researchers to fill gaps in these areas have been identified. Furthermore, Evaluation papers made up a clear majority of research approaches for studies explored (nearly half, 120 papers), followed by Proposed Solution and Solution with Validation. However, we noticed that there was a lack of validation for the solutions that are provided. When considering the types of contributions these papers made, a majority of papers contributed a solution or answer, or suggested a procedure or technique. These contributions were often in response to an identified problem by the authors. In the evaluation of *Programming/Development* papers, we observed that on average 58% of the quality questions identified were met, and there was a relationship between the length of research papers and the quality of such outcomes. This pattern does not auger well for the Stack Overflow community; however, recent trends suggest that there is likely to be an increase in validated outcomes. Such outcomes should also consider the delivery of mechanisms to improve the content that is provided on Stack Overflow, as against just reviewing what is provided.

This paper creates a starting point to understand the research that is dedicated to the study of Stack Overflow. In addition, outcomes from our investigations also identify specific avenues for future research. Furthermore, our research quality evaluations could direct replication studies, both in terms of the issues to consider and how studies should be designed. We believe that the classification schemes that were adapted in this work could be of utility for future software engineering researchers. In this regard, further research is required to validate solutions and contributions, and in particular, to explore how these may affect the validity of answers that are provided on Stack

Overflow. Such efforts may ensure that Stack Overflow remains a relevant, up-to-date and high quality CQA platform.

# Appendix A: Final List of Included Papers

| ID | Title | Year |
|---|---|---|
| P1 | Source code retrieval on StackOverflow using LDA | 2015 |
| P2 | Building reputation in StackOverflow: An empirical investigation | 2013 |
| P3 | On the Dynamics of Topic-Based Communities in Online Knowledge-Sharing Networks | 2015 |
| P4 | Badges of Friendship: Social Influence and Badge Acquisition on Stack Overflow | 2014 |
| P5 | A discriminative model approach for suggesting tags automatically for Stack Overflow questions | 2013 |
| P6 | Crowdsourced bug triaging | 2015 |
| P7 | NIRMAL: Automatic identification of software relevant tweets leveraging language model | 2015 |
| P8 | An effective experts mining technique in online discussion forums | 2016 |
| P9 | Editing Unfit Questions in Q&A | 2016 |
| P10 | Persuasive patterns in Q&A social networks | 2016 |
| P11 | Automated API Documentation with Tutorials Generated From Stack Overflow | 2016 |
| P12 | The role of comments' controversy in large-scale online discussion forums | 2016 |
| P13 | Harnessing Stack Overflow for the IDE | 2012 |
| P14 | Among the Machines: Human-Bot Interaction on Social Q&A Websites | 2016 |
| P15 | User Profiling for Answer Quality Assessment in Q&A Communities | 2013 |
| P16 | How the R Community Creates and Curates Knowledge: A Comparative Study of Stack Overflow and Mailing Lists | 2016 |
| P17 | Involvement, Contribution and Influence in GitHub and Stack Overflow | 2014 |
| P18 | Investigations into the goodness of posts in Q&A Forums—Popularity versus quality | 2015 |
| P19 | Predicting post importance in question answer forums based on topic-wise user expertise | 2015 |
| P20 | Stack Overflow Badges and User Behavior: An Econometric Approach | 2015 |
| P21 | Native-2-native: Automated Cross-platform Code Synthesis from Web-based Programming Resources | 2015 |
| P22 | A machine learning approach to cluster the users of stack overflow forum | 2015 |
| P23 | Discovering Value from Community Activity on Focused Question Answering Sites: A Case Study of Stack Overflow | 2012 |
| P24 | Steering User Behavior with Badges | 2013 |
| P25 | On the Personality Traits of StackOverflow Users | 2013 |
| P26 | Analyzing the friendliness of exchanges in an online software developer community | 2013 |
| P27 | Gender, Representation and Online Participation: A Quantitative Study of StackOverflow | 2012 |
| P28 | StackOverflow and GitHub: Associations between Software Development and Crowdsourced Knowledge | 2013 |
| P29 | Exploring user expertise and descriptive ability in community question answering | 2014 |
| P30 | Tag-based expert recommendation in community question answering | 2014 |
| P31 | Predicting the quality of questions on stackoverflow | 2015 |
| P32 | What are developers talking about? An analysis of topics and trends in Stack Overflow | 2014 |
| P33 | Facilitating crowd sourced software engineering via stack overflow | 2013 |
| P34 | Information security in software engineering, analysis of developers communications about security in social Q&A website | 2016 |
| P35 | Modeling Problem Difficulty and Expertise in Stackoverflow | 2012 |
| P36 | A Weighted Question Retrieval Model Using Descriptive Information in Community Question Answering | 2016 |
| P37 | Effects of tag usage on question response time: Analysis and prediction in StackOverflow | 2015 |
| P38 | Recognizing Gender of Stack Overflow Users | 2016 |
| P39 | Predicting Semantically Linkable Knowledge in Developer Online Forums via Convolutional Neural Network | 2016 |
| P40 | Domain-specific Cross-language Relevant Question Retrieval | 2016 |
| P41 | Mining Analogical Libraries in Q&A Discussions -- Incorporating Relational and Categorical Knowledge into Word Embedding | 2016 |
| P42 | Towards Correlating Search on Google and Asking on Stack Overflow | 2016 |
| P43 | A study of innovation diffusion through link sharing on stack overflow | 2013 |
| P44 | Mining Stack Overflow for discovering error patterns in SQL queries | 2015 |
| P45 | Challenges in Analyzing Software Documentation in Portuguese | 2015 |
| P46 | Searching crowd knowledge to recommend solutions for API usage tasks | 2016 |
| P47 | CODES: Mining Source Code Descriptions from Developers Discussions | 2014 |
| P48 | Can gamification motivate voluntary contributions? The case of StackOverflow Q&A community | 2015 |
| P49 | Building a domain knowledge base from wikipedia: A semi-supervised approach | 2016 |
| P50 | Augmenting API Documentation with Insights from Stack Overflow | 2016 |
| P51 | How Do Programmers Ask and Answer Questions on the Web? (NIER Track) | 2011 |
| P52 | Answers or no answers: Studying question answerability in Stack Overflow | 2015 |
| P53 | Mining Technology Landscape from Stack Overflow | 2016 |
| P54 | SimilarTech: Automatically Recommend Analogical Libraries Across Different Programming Languages | 2016 |
| P55 | Chaff from the wheat : Characterization and modeling of deleted questions on Stack Overflow | 2014 |

| ID | Title | Year |
|---|---|---|
| P56 | Determining the popularity of design patterns used by programmers based on the analysis of questions and answers on stackoverflow.Com social network | 2016 |
| P57 | Integrating Issue Tracking Systems with Community-Based Question and Answering Websites | 2013 |
| P58 | Recognizing gender differences in stack overflow usage: Applying the Bechdel test | 2016 |
| P59 | Towards a Weighted Voting System for Q&A Sites | 2013 |
| P60 | Software-Specific Named Entity Recognition in Software Engineering Social Content | 2016 |
| P61 | Analysis of the Reputation System and User Contributions on a Question Answering Website: StackOverflow | 2013 |
| P62 | Exploiting User Feedback to Learn to Rank Answers in Q&a Forums: A Case Study with Stack Overflow | 2013 |
| P63 | Extracting Skill Endorsements from Personal Communication Data | 2016 |
| P64 | Early Detection of Topical Expertise in Community Question Answering | 2015 |
| P65 | Partially Labeled Supervised Topic Models for Retrieving Similar Questions in CQA Forums | 2015 |
| P66 | Software-specific Part-of-speech Tagging: An Experimental Study on Stack Overflow | 2016 |
| P67 | Paradise Unplugged: Identifying Barriers for Female Participation on Stack Overflow | 2016 |
| P68 | Geo-locating the Knowledge Transfer in StackOverflow | 2013 |
| P69 | Fit or Unfit: Analysis and Prediction of 'Closed Questions' on Stack Overflow | 2013 |
| P70 | From Query to Usable Code: An Analysis of Stack Overflow Code Snippets | 2016 |
| P71 | What do practitioners ask about code clone? a preliminary investigation of stack overflow | 2015 |
| P72 | Identifying Developers' Expertise in Social Coding Platforms | 2016 |
| P73 | Automatic Categorization of Questions from Q&A Sites | 2014 |
| P74 | Contributor Motivation in Online Knowledge Sharing Communities with Reputation Management Systems | 2015 |
| P75 | Mining Successful Answers in Stack Overflow | 2015 |
| P76 | Moving to Stack Overflow: Best-Answer Prediction in Legacy Developer Forums | 2016 |
| P77 | Finding Expert Users in Community Question Answering | 2012 |
| P78 | Crowd Debugging | 2015 |
| P79 | Fundamentals of a graph transformation based web data processing system | 2014 |
| P80 | Automatically augmenting learning material with practical questions to increase its relevance | 2015 |
| P81 | Naturalness of Natural Language Artifacts in Software | 2015 |
| P82 | linking accounts across social networks: the case of StackOverflow, Github and twitter | 2015 |
| P83 | Topic shifts in stackoverflow: Ask it like socrates | 2016 |
| P84 | Learning a Dual-language Vector Space for Domain-specific Cross-lingual Question Retrieval | 2016 |
| P85 | Mining Questions About Software Energy Consumption | 2014 |
| P86 | A Study on the Most Popular Questions About Concurrent Programming | 2015 |
| P87 | Predicting Best Answerers for New Questions: An Approach Leveraging Distributed Representations of Words in Community Question Answering | 2015 |
| P88 | A Hybrid Model for Experts Finding in Community Question Answering | 2015 |
| P89 | Multistaging to understand: Distilling the essence of java code examples | 2016 |
| P90 | Linking Issue Tracker with Q&A Sites for Knowledge Sharing across Communities | 2015 |
| P91 | Predicting Questions' Scores on Stack Overflow | 2016 |
| P92 | Going Green: An Exploratory Analysis of Energy-related Questions | 2015 |
| P93 | Facing Up to the Inequality of Crowdsourced API Documentation | 2012 |
| P94 | CPDScorer: Modeling and evaluating developer programming ability across software communities | 2016 |
| P95 | Source Code Curation on StackOverflow: The Vesperin System | 2015 |
| P96 | Predicting Churn of Expert Respondents in Social Networks Using Data Mining Techniques: A Case Study of Stack Overflow | 2015 |
| P97 | Why is Stack Overflow Failing? Preserving Sustainability in Community Question Answering | 2016 |
| P98 | Modelling User Collaboration in Social Networks Using Edits and Comments | 2016 |
| P99 | Diagnosis at scale: Detecting the expertise level and knowledge states of lifelong professional learners | 2016 |
| P100 | Utilizing Non-QA Data to Improve Questions Routing for Users with Low QA Activity in CQA | 2015 |
| P101 | A Study of Demand-Driven Documentation in Two Open Source Projects | 2015 |
| P102 | Deficient documentation detection a methodology to locate deficient project documentation using topic analysis | 2013 |
| P103 | Data analysis of social community reputation: Good questions vs. good answers | 2015 |
| P104 | Multi-class multi-tag classifier system for StackOverflow questions | 2015 |
| P105 | Mining Web Technical Discussions to Identify Malware Capabilities | 2013 |
| P106 | ETA: Estimated Time of Answer Predicting Response Time in Stack Overflow | 2015 |
| P107 | Asking the Right Question in Collaborative Q&a Systems | 2014 |
| P108 | An Empirical Study on Stack Overflow Using Topic Analysis | 2015 |
| P109 | From Discussion to Wisdom: Web Resource Recommendation for Hyperlinks in Stack Overflow | 2016 |
| P110 | API Deprecation: A Retrospective Analysis and Detection Method for Code Examples on the Web | 2016 |
| P111 | Automatic mapping of user tags to Wikipedia concepts: The case of a Q&A website - StackOverflow | 2015 |

| ID | Title | Year |
|---|---|---|
| P112 | Text mining stackoverflow: An insight into challenges and subject-related difficulties faced by computer science learners | 2016 |
| P113 | Mining Questions Asked by Web Developers | 2014 |
| P114 | Code search with input/output queries: Generalizing, ranking, and assessment | 2016 |
| P115 | Using and asking: APIs used in the Android market and asked about in StackOverflow | 2013 |
| P116 | Software trustworthiness 2.0 - A semantic web enabled global source code analysis approach | 2014 |
| P117 | Perceptions of Answer Quality in an Online Technical Question and Answer Forum | 2014 |
| P118 | On the Extraction of Cookbooks for APIs from the Crowd Knowledge | 2014 |
| P119 | The influence of App churn on App success and StackOverflow discussions | 2015 |
| P120 | Query Expansion Based on Crowd Knowledge for Code Search | 2016 |
| P121 | Leveraging Crowd Knowledge for Software Comprehension and Development | 2013 |
| P122 | Understanding and Classifying the Quality of Technical Forum Questions | 2014 |
| P123 | Improving Low Quality Stack Overflow Post Detection | 2014 |
| P124 | Prompter: A Self-Confident Recommender System | 2014 |
| P125 | Leveraging Informal Documentation to Summarize Classes and Methods in Context | 2015 |
| P126 | Retrieving rising stars in focused community question-answering | 2016 |
| P127 | Design Lessons from the Fastest Q&a Site in the West | 2011 |
| P128 | Off the Beaten Path: Let's Replace Term-Based Retrieval with k-NN Search | 2016 |
| P129 | Predict closed questions on StackOverflow | 2013 |
| P130 | Is It good to be like wikipedia?: Exploring the trade-offs of introducing collaborative editing model to QandA Sites | 2015 |
| P131 | Quantifying the impact of badges on user engagement in online Q&A communities | 2012 |
| P132 | CQArank: Jointly Model Topics and Expertise in Community Question Answering | 2013 |
| P133 | Incorporating social information to perform diverse replier recommendation in question and answer communities | 2016 |
| P134 | Interruptible Tasks: Treating Memory Pressure As Interrupts for Highly Scalable Data-parallel Programs | 2015 |
| P135 | Seahawk: Stack Overflow in the IDE | 2013 |
| P136 | StORMeD: Stack Overflow Ready Made Data | 2015 |
| P137 | CodeTube: Extracting Relevant Fragments from Software Development Video Tutorials | 2016 |
| P138 | Too Long; Didn't Watch! Extracting Relevant Fragments from Software Development Video Tutorials | 2016 |
| P139 | Mining StackOverflow to Turn the IDE into a Self-confident Programming Prompter | 2014 |
| P140 | A Task Decomposition Framework for Surveying the Crowd Contextual Insights | 2015 |
| P141 | Harnessing Implicit Teamwork Knowledge to Improve Quality in Crowdsourcing Processes | 2014 |
| P142 | Why, when, and what: Analyzing Stack Overflow questions by topic, type, and code | 2013 |
| P143 | Answering questions about unanswered questions of Stack Overflow | 2013 |
| P144 | Who Will Answer My Question on Stack Overflow? | 2015 |
| P145 | An exploratory analysis of mobile development issues using stack overflow | 2013 |
| P146 | RACK: Automatic API Recommendation Using Crowdsourced Knowledge | 2016 |
| P147 | Recommending insightful comments for source code using crowdsourced knowledge | 2015 |
| P148 | An IDE-based context-aware meta search engine | 2013 |
| P149 | Towards a context-aware IDE-based meta search engine for recommendation about programming errors and exceptions | 2014 |
| P150 | An Empirical Study on the Usage of the Swift Programming Language | 2016 |
| P151 | Architectural Knowledge for Technology Decisions in Developer Communities: An Exploratory Study with StackOverflow | 2016 |
| P152 | "Should We Move to Stack Overflow?" Measuring the Utility of Social Media for Developer Support | 2015 |
| P153 | "A Bit of Code": How the Stack Overflow Community Creates Quality Postings | 2014 |
| P154 | A tri-role topic model for domain-specific question answering | 2015 |
| P155 | Quality Questions Need Quality Code: Classifying Code Fragments on Stack Overflow | 2015 |
| P156 | Static Analysis of Event-driven Node.Js JavaScript Applications | 2015 |
| P157 | A system for scalable and reliable technical-skill testing in online labor markets | 2015 |
| P158 | How Do API Changes Trigger Stack Overflow Discussions? A Study on the Android SDK | 2014 |
| P159 | Mining and Comparing Engagement Dynamics Across Multiple Social Media Platforms | 2014 |
| P160 | Knowledge fixation and accretion: Longitudinal analysis of a social question-answering site | 2014 |
| P161 | Leveraging a Corpus of Natural Language Descriptions for Program Similarity | 2016 |
| P162 | Detecting topics and overlapping communities in question and answer sites | 2015 |
| P163 | Requirement Acquisition from Social Q&A Sites | 2015 |
| P164 | QUICKAR: Automatic Query Reformulation for Concept Location Using Crowdsourced Knowledge | 2016 |
| P165 | An Insight into the Unresolved Questions at Stack Overflow | 2015 |
| P166 | Improving the Quality of Code Snippets in Stack Overflow | 2016 |
| P167 | Embedded emotion-based classification of stack overflow questions towards the question quality prediction | 2016 |
| P168 | Mining Duplicate Questions in Stack Overflow | 2016 |

| ID | Title | Year |
| --- | --- | --- |
| P169 | The Synergy between Voting and Acceptance of Answers on StackOverflow - Or the Lack Thereof | 2015 |
| P170 | SODA: the stack overflow dataset almanac | 2015 |
| P171 | The Challenges of Sentiment Detection in the Social Programmer Ecosystem | 2015 |
| P172 | Towards Discovering the Role of Emotions in Stack Overflow | 2014 |
| P173 | Participation Differences in Q&A Sites Across Countries: Opportunities for Cultural Adaptation | 2016 |
| P174 | Distribution, correlation and prediction of response times in Stack Overflow | 2014 |
| P175 | What Do Client Developers Concern When Using Web APIs? An Empirical Study on Developer Forums and Stack Overflow | 2016 |
| P176 | Is programming knowledge related to age? An exploration of stack overflow | 2013 |
| P177 | Veteran developers' contributions and motivations: An open source perspective | 2016 |
| P178 | Mining Testing Questions on Stack Overflow | 2016 |
| P179 | Discovering Essential Code Elements in Informal Documentation | 2013 |
| P180 | What programmers say about refactoring tools? An empirical investigation of stack overflow | 2013 |
| P181 | The Good, the Bad and Their Kins: Identifying Questions with Negative Scores in StackOverflow | 2015 |
| P182 | Nearest Neighbour Based Transformation Functions for Text Classification: A Case Study with StackOverflow | 2016 |
| P183 | Prompter: Turning the IDE into a self-confident programming assistant | 2016 |
| P184 | Using Semantics to Search Answers for Unanswered Questions in Q&A Forums | 2016 |
| P185 | Linked Data in Crowdsourcing Purposive Social Network | 2013 |
| P186 | User churn in focused question answering sites: Characterizations and prediction | 2014 |
| P187 | Ranking open source software based on crowd wisdom | 2015 |
| P188 | Choosing your weapons: On sentiment analysis tools for software engineering research | 2015 |
| P189 | Is Stack Overflow Overflowing With Questions and Tags | 2015 |
| P190 | One-Day Flies on StackOverflow - Why the Vast Majority of StackOverflow Users Only Posts Once | 2015 |
| P191 | Asking for (and about) permissions used by Android apps | 2013 |
| P192 | Towards Improving Bug Tracking Systems with Game Mechanisms | 2012 |
| P193 | Footprint Model for Discussion Forums in MOOC | 2015 |
| P194 | Discovery of Technical Expertise from Open Source Code Repositories | 2013 |
| P195 | Mining community-based top-k experts and learners in online question answering systems | 2016 |
| P196 | Toward Understanding the Causes of Unanswered Questions in Software Information Sites: A Case Study of Stack Overflow | 2013 |
| P197 | Learning from Gurus: Analysis and Modeling of Reopened Questions on Stack Overflow | 2016 |
| P198 | What are mobile developers asking about? A large scale study using stack overflow | 2016 |
| P199 | A Manual Categorization of Android App Development Issues on Stack Overflow | 2014 |
| P200 | Automatic Assessments of Code Explanations: Predicting Answering Times on Stack Overflow | 2015 |
| P201 | Encouraging user behaviour with achievements: An empirical study | 2013 |
| P202 | What makes a good code example?: A study of programming Q&A in StackOverflow | 2012 |
| P203 | Entity disambiguation in tweets leveraging user social profiles | 2013 |
| P204 | Making sense of online code snippets | 2013 |
| P205 | EnTagRec: An Enhanced Tag Recommendation System for Software Information Sites | 2014 |
| P206 | An empirical analysis on reducing open source software development tasks using stack overflow | 2016 |
| P207 | Crowdsourced bug triaging: Leveraging Q&A platforms for bug assignment | 2016 |
| P208 | Purposes, Concepts, Misfits, and a Redesign of Git | 2016 |
| P209 | Jumping Through Hoops: Why Do Java Developers Struggle with Cryptography APIs? | 2016 |
| P210 | Predicting Answering Times on Stack Overflow | 2015 |
| P211 | Security Expert Recommender in Software Engineering | 2016 |
| P212 | Mining StackOverflow to Filter out Off-topic IRC Discussion | 2015 |
| P213 | An Empirical Study on Developer Interactions in StackOverflow | 2013 |
| P214 | DIETs: Recommender Systems for Mobile API Developers | 2015 |
| P215 | Grouping Android Tag Synonyms on Stack Overflow | 2016 |
| P216 | Synonym Suggestion for Tags on Stack Overflow | 2015 |
| P217 | Multivariate Beta Mixture Model for Automatic Identification of Topical Authoritative Users in Community Question Answering Sites | 2016 |
| P218 | Selecting Best Answer: An Empirical Analysis on Community Question Answering Sites | 2016 |
| P219 | Automatic Knowledge Sharing Across Communities: A Case Study on Android Issue Tracker and Stack Overflow | 2015 |
| P220 | Linking Stack Overflow to Issue Tracker for Issue Resolution | 2014 |
| P221 | T2API: Synthesizing API Code Usage Templates from English Texts with Statistical Translation | 2016 |
| P222 | Employing Source Code Information to Improve Question-answering in Stack Overflow | 2015 |
| P223 | Towards predicting the best answers in community-based question-answering services | 2013 |
| P224 | Predicting best answerers for new questions: An approach leveraging topic modeling and collaborative voting | 2014 |
| P225 | Interrogative-guided Re-ranking for Question-oriented Software Text Retrieval | 2014 |

| ID | Title | Year |
|---|---|---|
| P226 | Topical Authoritative Answerer Identification on Q&A Posts Using Supervised Learning in CQA Sites | 2016 |
| P227 | An Empirical Analysis of a Network of Expertise | 2013 |
| P228 | Min(e)d your tags: Analysis of Question response time in StackOverflow | 2014 |
| P229 | A Hybrid Auto-tagging System for StackOverflow Forum Questions | 2014 |
| P230 | Understanding the Usage of Online Forums as Learning Platforms | 2015 |
| P231 | ExceptionTracer: A Solution Recommender for Exceptions in an Integrated Development Environment | 2015 |
| P232 | CSNIPPEX: Automated Synthesis of Compilable Code Snippets from Q&A Sites | 2016 |
| P233 | Structurally Heterogeneous Source Code Examples from Unstructured Knowledge Sources | 2015 |
| P234 | Intuition vs. Truth: Evaluation of Common Myths About Stackoverflow Posts | 2015 |
| P235 | TBIL: A Tagging-Based Approach to Identity Linkage Across Software Communities | 2015 |
| P236 | SOLinker: Constructing Semantic Links between Tags and URLs on StackOverflow | 2016 |
| P237 | Recommending Posts concerning API Issues in Developer Q&A Sites | 2015 |
| P238 | Detecting API usage obstacles: A study of iOS and Android developer questions | 2013 |
| P239 | How do developers react to RESTful API evolution? | 2014 |
| P240 | Search literacy: Learning to search to learn | 2016 |
| P241 | GRETA: Graph-Based Tag Assignment for GitHub Repositories | 2016 |
| P242 | Exploiting User Feedback for Expert Finding in Community Question Answering | 2015 |
| P243 | A Hybrid Trust-Based Recommender System for Online Communities of Practice | 2015 |
| P244 | It Takes Two to Tango: Deleted Stack Overflow Question Prediction with Text and Meta Features | 2016 |
| P245 | Tag recommendation in software information sites | 2013 |
| P246 | Relationship-aware Code Search for JavaScript Frameworks | 2016 |
| P247 | You Get Where You're Looking for: The Impact of Information Sources on Code Security | 2016 |
| P248 | Understanding Sequential User Behavior in Social Computing: To Answer or to Vote? | 2015 |
| P249 | Automated construction of a software-specific word similarity database | 2014 |
| P250 | Evaluating Bug Severity Using Crowd-based Knowledge: An Exploratory Study | 2015 |
| P251 | E-WISE: An expertise-driven recommendation platform for Web question answering systems | 2015 |
| P252 | Harnessing engagement for knowledge creation acceleration in collaborative Q&A systems | 2015 |
| P253 | Sparrows and owls: Characterisation of expert behaviour in StackOverflow | 2014 |
| P254 | What Security Questions Do Developers Ask? A Large-Scale Study of Stack Overflow Posts | 2016 |
| P255 | CUT: A combined approach for tag recommendation in software information sites | 2016 |
| P256 | The structure and dynamics of knowledge network in domain-specific Q&A sites: a case study of stack overflow | 2016 |
| P257 | Quick Trigger on Stack Overflow: A Study of Gamification-influenced Member Tendencies | 2015 |
| P258 | SEWordSim: Software-specific Word Similarity Database | 2014 |
| P259 | Simplified Detection and Labeling of Overlapping Communities of Interest in Question-and-Answer Sites | 2015 |
| P260 | Empirical study on overlapping community detection in question and answer sites | 2014 |
| P261 | Multi-Factor Duplicate Question Detection in Stack Overflow | 2015 |
| P262 | Crowd-selection query processing in crowdsourcing databases: A task-driven approach | 2015 |
| P263 | Building a large-scale software programming taxonomy from stackoverflow | 2015 |
| P264 | Learning to Rank for Question Routing in Community Question Answering | 2013 |
| P265 | Learning to rank for question-oriented software text retrieval | 2015 |

## Appendix B: Research Study Quality Evaluation Dimensions[10]

A. Are the aims of the study stated clearly?
B. Is the basis of evaluative appraisal clear?
   - Discussion of how assessments of effectiveness/evaluative judgements have been reached *(i.e. whose judgements are they and on what basis have they been reached?)*
   - Description of any formalised appraisal criteria used, when generated and how and by whom they have been applied
   - Discussion of the nature and source of any divergence in evaluative appraisals
   - Discussion of any unintended consequences of intervention, their impact and why they arose
C. How defensible is the research design?
   - Discussion of how overall research strategy was designed to meet aims of study
   - Discussion of rationale for study design
   - Convincing argument for different features of research design (e.g. reasons given for different components or stages of research; purpose of particular methods or data sources, multiple methods, time frames etc.)
   - Use of different features of design/data sources evident in findings presented
   - Discussion of limitations of research design and their implications for the study evidence
D. Are data collection methods described adequately?
   - "The researchers should define all key variables and provide information on the type of measure (e.g. test, survey), its content, and its length. If a measure is administered more than once (e.g. before and after) an intervention, check to see that the length of time between administrations is explained and that its potential effect on test-retest reliability is discussed."[11]
E. Has the approach to, and formulation of, analysis been conveyed adequately?
   - Description of form of original data (e.g. use of verbatim transcripts, observation or interview notes, documents, etc.)
   - Clear rationale for choice of data management method/tool/package
   - Evidence of how descriptive analytic categories, classes, labels etc. have been generated and used (i.e. either through explicit discussion or portrayal in the commentary)
   - Discussion, with examples, of how any constructed analytic concepts/typologies etc. have been devised and applied
F. Has the diversity of perspectives and contexts been explored?
   - Discussion of contribution of sample design/ case selection in generating diversity
   - Description and illumination of diversity/multiple perspectives/alternative positions in the evidence displayed
   - Evidence of attention to negative cases, outliers or exceptions
   - Typologies/models of variation derived and discussed
   - Examination of origins/influences on opposing or differing positions
   - Identification of patterns of association/linkages with divergent positions/groups
G. Are there any links between data, interpretation and conclusions?

---

[10] Kitchenham, B., & Charters, S. (2007). Guidelines for performing systematic literature reviews in software engineering *Technical report, Ver. 2.3 EBSE Technical Report. EBSE*. and Spencer, L., Ritchie, J., Lewis, J., & Dillon, L. (2003). Quality in qualitative evaluation: a framework for assessing research evidence.

[11] Fink, A. (2010). *Conducting research literature reviews: From the internet to paper* (3rd ed.). Los Angeles: Sage.

- o Clear conceptual links between analytic commentary and presentations of original data (i.e. commentary and cited data relate; there is an analytic context to cited data, not simply repeated description)
- o Discussion of how/why particular interpretation/significance is assigned to specific aspects of data – with illustrative extracts of original data
- o Discussion of how explanations/ theories/conclusions were derived – and how they relate to interpretations and content of original data (i.e. how warranted); whether alternative explanations explored
- o Display of negative cases and how they lie outside main proposition/theory/ hypothesis etc.; or how proposition etc. revised to include them

H. Is the reporting clear and coherent?
- o Demonstrates link to aims of study/research questions
- o Provides a narrative/story or clearly constructed thematic account
- o Has structure and signposting that usefully guide reader through the commentary
- o Provides accessible information for intended target audience(s)
- o Key messages highlighted or summarised

I. Has the research process been documented adequately?
- o Discussion of strengths and weaknesses of data sources and methods
- o Documentation of changes made to design and reasons; implications for study coverage
- o Documentation and reasons for changes in sample coverage/data collection/analytic approach; implications
- o Reproduction of main study documents (e.g. letters of approach, topic guides, observation templates, data management frameworks etc.)

J. Could the study be replicated?

# Appendix C: Full List of Venues

| Venue | # of papers |
|---|---|
| IEEE Working Conference on Mining Software Repositories | 40 |
| IEEE International Conference on Software Engineering | 13 |
| Lecture Notes in Computer Science | 13 |
| IEEE International Conference on Software Maintenance | 10 |
| IEEE/ACM International Conference on Advances in Social Networks Analysis and Mining | 8 |
| ACM SIGSOFT International Symposium on Foundations of Software Engineering | 6 |
| ACM/IEEE International Conference on Automated Software Engineering | 6 |
| International Conference on Software Analysis, Evolution, and Reengineering | 6 |
| International Conference on World Wide Web | 6 |
| Annual ACM Symposium on Applied Computing | 5 |
| IEEE International Conference on Program Comprehension | 5 |
| ACM International Conference on Information & Knowledge Management | 4 |
| Annual Computer Software and Applications Conference | 4 |
| CEUR Workshop Proceedings | 4 |
| Empirical Software Engineering | 4 |
| International Conference on Software Engineering and Knowledge Engineering | 4 |
| Advances in Intelligent Systems and Computing | 3 |
| Brazilian Symposium on Software Engineering | 3 |
| Hawaii International Conference on System Sciences | 3 |
| International Workshop on Social Software Engineering | 3 |
| Journal of Information Science | 3 |
| ACM Conference on Computer Supported Cooperative Work | 2 |
| IEEE Conference on Software Maintenance, Reengineering and Reverse Engineering | 2 |
| ACM Conference on Hypertext and Social Media | 2 |
| ACM International Conference on the Theory of Information Retrieval | 2 |
| ACM SIGPLAN International Conference on Object-Oriented Programming, Systems, Languages, and Applications | 2 |
| ACM/IEEE International Symposium on Empirical Software Engineering and Measurement | 2 |
| Asia-Pacific Symposium on Internetware | 2 |
| Australasian Software Engineering Conference | 2 |
| CHI Conference Extended Abstracts on Human Factors in Computing Systems | 2 |
| IEEE Access | 2 |
| IEEE Symposium on Visual Languages and Human-Centric Computing | 2 |
| IEEE Transactions on Services Computing | 2 |
| International ACM SIGIR Conference on Research and Development in Information Retrieval | 2 |
| International Conference on Service-Oriented Computing and Applications | 2 |
| International Workshop on Cooperative and Human Aspects of Software Engineering | 2 |
| Journal of Computer Science and Technology | 2 |
| Journal of Systems and Software | 2 |
| Procedia Computer Science | 2 |
| Social Network Analysis and Mining | 2 |
| ACM Conference on Online Social Networks | 1 |
| ACM Conference on Web Science | 1 |
| ACM International Conference on Computer-Supported Cooperative Work and Social Computing | 1 |
| ACM International Symposium on New Ideas, New Paradigms, and Reflections on Programming and Software | 1 |
| ACM SIGKDD International Conference on Knowledge Discovery and Data Mining | 1 |
| ACM SIGPLAN International Conference on Generative Programming: Concepts and Experiences | 1 |
| ACM Workshop on Refactoring Tools | 1 |
| Annual ACM India Conference | 1 |
| Annual International Conference on Computer Science and Software Engineering | 1 |
| Annual Research Conference on South African Institute of Computer Scientists and Information Technologists | 1 |
| Asia-Pacific Software Engineering Conference | 1 |
| Communications in Computer and Information Science | 1 |
| Computer Networks | 1 |
| Conference on User Modeling Adaptation and Personalization | 1 |
| Euromicro Conference on Software Engineering and Advanced Applications | 1 |
| European Conference on Software Maintenance and Reengineering | 1 |
| Finding Source Code on the Web for Remix and Reuse | 1 |

| Venue | # of papers |
| --- | --- |
| Frontiers in Education Conference | 1 |
| IEEE Conference on Cognitive Infocommunications | 1 |
| IEEE International Autumn Meeting on Power, Electronics and Computing | 1 |
| IEEE International Conference on Data Mining Workshop | 1 |
| IEEE International Conference on Industrial Engineering and Engineering Management | 1 |
| IEEE International Conference on Software Engineering and Service Science | 1 |
| IEEE International Conference on Web Services | 1 |
| IEEE Software | 1 |
| IEEE Symposium on Security and Privacy | 1 |
| IEEE Symposium on Service-Oriented System Engineering | 1 |
| IEEE Transactions on Learning Technologies | 1 |
| IEEE Transactions on Network Science and Engineering | 1 |
| IEEE/WIC/ACM International Conference on Web Intelligence and Intelligent Agent Technology | 1 |
| IKDD Conference on Data Science | 1 |
| India Software Engineering Conference | 1 |
| Indian Journal of Science and Technology | 1 |
| International Conference Companion on World Wide Web | 1 |
| International Conference on Collaborative Computing: Networking, Applications and Worksharing | 1 |
| International Conference on Computing, Electronic and Electrical Engineering | 1 |
| International Conference on Cyber-Enabled Distributed Computing and Knowledge Discovery | 1 |
| International Conference on Distributed Computing Systems Workshops | 1 |
| International Conference on Extending Database Technology | 1 |
| International Conference on Frontier of Computer Science and Technology | 1 |
| International Conference on Information and Communication Technology | 1 |
| International Conference on Information Reuse and Integration | 1 |
| International Conference on Information Systems | 1 |
| International Conference on Interdisciplinary Advances in Applied Computing | 1 |
| International Conference on Machine Learning and Applications | 1 |
| International Conference on Quality Software | 1 |
| International Conference on Research in Adaptive and Convergent Systems | 1 |
| International Conference on Social Computing | 1 |
| International Conference on Social Informatics | 1 |
| International Conference on Weblogs and Social Media | 1 |
| International Conference Recent Advances in Natural Language Processing | 1 |
| International Congress on Advanced Applied Informatics | 1 |
| International Symposium on Software Testing and Analysis | 1 |
| International Symposium on Women in Computing and Informatics | 1 |
| International Working Conference on Source Code Analysis and Manipulation | 1 |
| International Workshop on CrowdSourcing in Software Engineering | 1 |
| International Workshop on Recommendation Systems for Software Engineering | 1 |
| International Workshop on Software Clones | 1 |
| International Workshop on Software Mining | 1 |
| Journal of Documentation | 1 |
| Journal of Enterprise Information Management | 1 |
| Journal of Software: Evolution and Process | 1 |
| National Conference on Artificial Intelligence | 1 |
| Nordic Conference on Human-Computer Interaction | 1 |
| Requirements Engineering in the Big Data Era | 1 |
| European Network Intelligence Conference | 1 |
| SIGSOFT Software Engineering Notes | 1 |
| Symposium on Operating Systems Principles | 1 |
| Working Conference on Reverse Engineering | 1 |
| Working IEEE/IFIP Conference on Software Architecture | 1 |
| Workshop on Data-driven User Behavioral Modelling and Mining from Social Media | 1 |
| Workshop on Evaluation and Usability of Programming Languages and Tools | 1 |
| Workshop on Mining Unstructured Data | 1 |
| Workshop on Partial Evaluation and Program Manipulation | 1 |

**Appendix D: Detailed Quality Outcomes for 58 Programming/Development Papers (ordered as they were analysed)**

| ID | # of Pages | A | B | C | D | E | F | G | H | I | J | Total |
|---|---|---|---|---|---|---|---|---|---|---|---|---|
| P153 | 10 | 1 | 0 | 1 | 1 | 1 | 1 | 1 | 1 | 1 | 1 | 9 |
| P199 | 5 | 1 | 0 | 0 | 1 | 0 | 0 | 0 | 0 | 0 | 1 | 3 |
| P86 | 8 | 1 | 0 | 1 | 1 | 1 | 1 | 1 | 1 | 1 | 1 | 9 |
| P108 | 4 | 1 | 1 | 0 | 1 | 1 | 0 | 0 | 0 | 0 | 0 | 4 |
| P150 | 5 | 1 | 0 | 1 | 1 | 1 | 1 | 1 | 1 | 0 | 1 | 8 |
| P145 | 4 | 1 | 0 | 0 | 1 | 1 | 0 | 0 | 1 | 0 | 1 | 5 |
| P148 | 5 | 1 | 1 | 0 | 0 | 0 | 0 | 0 | 1 | 0 | 1 | 4 |
| P191 | 10 | 1 | 0 | 1 | 0 | 1 | 1 | 1 | 1 | 1 | 1 | 8 |
| P200 | 4 | 1 | 0 | 0 | 0 | 1 | 0 | 1 | 0 | 0 | 0 | 3 |
| P263 | 6 | 1 | 1 | 1 | 1 | 1 | 0 | 0 | 1 | 1 | 1 | 8 |
| P114 | 14 | 1 | 1 | 1 | 1 | 1 | 1 | 1 | 1 | 1 | 1 | 10 |
| P47 | 4 | 0 | 0 | 0 | 1 | 0 | 0 | 0 | 0 | 0 | 0 | 1 |
| P137 | 4 | 0 | 0 | 0 | 0 | 1 | 0 | 0 | 1 | 0 | 0 | 2 |
| P78 | 13 | 1 | 1 | 1 | 1 | 1 | 0 | 1 | 1 | 1 | 1 | 9 |
| P232 | 12 | 1 | 1 | 1 | 1 | 1 | 1 | 1 | 1 | 1 | 1 | 10 |
| P56 | 13 | 0 | 0 | 0 | 0 | 0 | 0 | 1 | 0 | 0 | 0 | 1 |
| P179 | 10 | 1 | 1 | 1 | 0 | 1 | 1 | 1 | 1 | 1 | 1 | 9 |
| P250 | 4 | 1 | 0 | 0 | 1 | 1 | 1 | 0 | 1 | 0 | 1 | 6 |
| P231 | 4 | 1 | 0 | 0 | 0 | 0 | 0 | 0 | 1 | 0 | 0 | 2 |
| P33 | 20 | 1 | 0 | 0 | 1 | 1 | 0 | 1 | 1 | 1 | 1 | 7 |
| P70 | 12 | 1 | 1 | 1 | 1 | 1 | 0 | 1 | 1 | 1 | 1 | 9 |
| P92 | 4 | 1 | 0 | 0 | 1 | 1 | 0 | 0 | 0 | 0 | 0 | 3 |
| P13 | 5 | 1 | 0 | 0 | 0 | 0 | 1 | 0 | 1 | 0 | 0 | 3 |
| P166 | 6 | 0 | 1 | 0 | 0 | 0 | 0 | 0 | 1 | 0 | 1 | 3 |
| P34 | 10 | 1 | 0 | 0 | 0 | 1 | 0 | 1 | 1 | 0 | 1 | 5 |
| P57 | 9 | 1 | 0 | 1 | 1 | 1 | 1 | 0 | 1 | 0 | 1 | 7 |
| P121 | 10 | 0 | 1 | 0 | 1 | 0 | 1 | 1 | 1 | 0 | 0 | 5 |
| P125 | 4 | 0 | 1 | 1 | 0 | 0 | 0 | 0 | 1 | 0 | 0 | 3 |
| P204 | 4 | 1 | 0 | 1 | 0 | 1 | 0 | 0 | 0 | 0 | 1 | 4 |
| P41 | 11 | 1 | 1 | 1 | 0 | 1 | 1 | 1 | 1 | 1 | 1 | 9 |
| P85 | 10 | 1 | 1 | 0 | 1 | 1 | 1 | 1 | 1 | 1 | 1 | 9 |
| P113 | 10 | 1 | 0 | 1 | 1 | 1 | 1 | 1 | 1 | 1 | 1 | 9 |
| P44 | 5 | 1 | 1 | 1 | 1 | 0 | 0 | 0 | 1 | 0 | 1 | 6 |
| P139 | 10 | 1 | 1 | 1 | 1 | 0 | 0 | 1 | 1 | 0 | 1 | 7 |
| P178 | 7 | 1 | 0 | 1 | 0 | 1 | 0 | 0 | 1 | 0 | 1 | 5 |
| P105 | 5 | 1 | 0 | 0 | 0 | 1 | 0 | 0 | 0 | 0 | 1 | 3 |
| P89 | 10 | 1 | 1 | 0 | 1 | 0 | 0 | 1 | 0 | 1 | 1 | 6 |
| P21 | 10 | 0 | 1 | 0 | 1 | 1 | 0 | 1 | 0 | 1 | 1 | 6 |
| P124 | 4 | 1 | 1 | 0 | 0 | 0 | 0 | 0 | 1 | 0 | 0 | 3 |
| P183 | 42 | 1 | 1 | 1 | 1 | 1 | 1 | 1 | 1 | 1 | 1 | 10 |
| P155 | 4 | 1 | 0 | 1 | 1 | 1 | 0 | 1 | 1 | 1 | 1 | 8 |
| P147 | 10 | 1 | 1 | 0 | 1 | 1 | 0 | 0 | 1 | 1 | 1 | 7 |
| P246 | 12 | 1 | 1 | 1 | 0 | 1 | 1 | 1 | 1 | 1 | 1 | 9 |
| P163 | 11 | 1 | 0 | 1 | 1 | 1 | 1 | 0 | 1 | 0 | 1 | 7 |
| P135 | 4 | 1 | 0 | 0 | 0 | 0 | 0 | 0 | 1 | 0 | 0 | 2 |
| P54 | 6 | 1 | 0 | 0 | 0 | 0 | 0 | 0 | 1 | 0 | 0 | 2 |
| P116 | 18 | 1 | 0 | 1 | 0 | 1 | 0 | 1 | 1 | 0 | 0 | 5 |
| P95 | 4 | 1 | 0 | 0 | 0 | 1 | 0 | 0 | 1 | 0 | 0 | 3 |
| P1 | 5 | 0 | 0 | 0 | 0 | 1 | 0 | 0 | 0 | 0 | 0 | 1 |
| P156 | 15 | 1 | 1 | 1 | 0 | 1 | 0 | 1 | 0 | 1 | 1 | 7 |
| P233 | 6 | 1 | 1 | 1 | 0 | 1 | 1 | 0 | 1 | 0 | 1 | 7 |
| P149 | 10 | 1 | 0 | 1 | 1 | 1 | 1 | 0 | 0 | 1 | 1 | 7 |
| P192 | 10 | 1 | 0 | 1 | 1 | 1 | 0 | 1 | 1 | 0 | 1 | 7 |
| P198 | 32 | 1 | 0 | 1 | 1 | 1 | 0 | 1 | 1 | 1 | 1 | 8 |
| P71 | 2 | 0 | 0 | 0 | 1 | 0 | 0 | 0 | 1 | 0 | 0 | 2 |

| ID | # of Pages | A | B | C | D | E | F | G | H | I | J | Total |
|---|---|---|---|---|---|---|---|---|---|---|---|---|
| P202 | 10 | 1 | 0 | 0 | 1 | 1 | 1 | 1 | 1 | 0 | 1 | 7 |
| P180 | 4 | 1 | 0 | 0 | 1 | 0 | 1 | 0 | 1 | 0 | 1 | 5 |
| P254 | 15 | 1 | 0 | 1 | 1 | 1 | 0 | 0 | 1 | 1 | 1 | 7 |

The questions corresponding to the quality dimensions A to J that appear in column headers are provided below, with full classification scheme in Appendix B.

A. Are the aims of the study stated clearly?

B. Is the basis of evaluative appraisal clear?

C. How defensible is the research design?

D. Are data collection methods described adequately?

E. Has the approach to, and formulation of, analysis been conveyed adequately?

F. Has the diversity of perspectives and contexts been explored?

G. Are there any links between data, interpretation and conclusions?

H. Is the reporting clear and coherent?

I. Has the research process been documented adequately?

J. Could the study be replicated?